\documentclass[aps,prb,twocolumn,amsmath,amssymb,nofootinbib,superscriptaddress,floatfix,eqsecnum]{revtex4-1}

\usepackage{amsmath}
\usepackage{amssymb}
\usepackage{amsthm}
\usepackage[dvipdf]{color}
\usepackage{graphicx}
\usepackage{dcolumn}
\usepackage{bm}

\begin{document}
\title{A Generic Microscopic Theory for the Universality of TTLS Model Meissner-Berret Ratio in Low-Temperature Glasses}

  \author{Di Zhou$^{1}$, Anthony J. Leggett}

 \affiliation{
 Department of Physics,
  University of Illinois at Urbana-Champaign,
  1110 West Green St, Urbana,
   Illinois 61801, USA
 }

\date{\today}

\begin{abstract}
Tunneling-two-level-system (TTLS) model has successfully explained several low-temperature glass universal properties which do not exist in their crystalline counterparts. The coupling constants between longitudinal and transverse phonon strain fields and two-level-systems are denoted as $\gamma_l$ and $\gamma_t$. The ratio $\gamma_l/\gamma_t$ was observed to lie between $1.44$ and $1.84$ for 18 different kinds of glasses. Such universal property cannot be explained within TTLS model. In this paper by developing a microscopic generic coupled block model, we show that the ratio $\gamma_l/\gamma_t$ is proportinal to the ratio of sound velocity $c_l/c_t$. We prove that the universality of $\gamma_l/\gamma_t$ essentially comes from the mutual interaction between different glass blocks, independent of the microscopic structure and chemical compound of the amorphous materials. In the appendix we also give a detailed correction on the coefficient of non-elastic stress-stress interaction $\Lambda_{ijkl}^{(ss')}$ which was obtained by Joffrin and Levelut\cite{Joffrin1976}.
\end{abstract}

\maketitle

\section{Introduction}
It has been more than 50 years since the first experiment\cite{Zeller1971} by Zeller and Pohl showed that at ultra-low temperatures below 1K, the thermal and acoustic properties of amorphous solids (glasses) behave strikingly different from that of the crystalline counterparts. Anderson, Halperin and Varma's\cite{Anderson1972} group and Phillips\cite{Phillips1987} independently developed a microscopic phenomenological model which was later known as the tunneling-two-level-system (TTLS) model, to explain the anomalous properties of low-temperature glasses. It successfully explained several universal properties of glasses which were never observed in crystalline solids, e.g., linear heat capacity (two orders of magnitude greater than the contribution of heat capacity from conducting electrons), saturation, echoes, quadratic heat conductivity etc..

In TTLS model, people assume that there are a group of tunneling-two-level-systems randomly embedded in the glass materials. The effective Hamiltonian of low-temperature glass in TTLS model is the summation of long wavelength phonon Hamiltonian, a set of (non-elastic) two-level-system Hamiltonian and the coupling between phonon strain fields and two-level-systems. The coupling constant between longitudinal phonon and two-level-system is denoted as $\gamma_l$; for transverse phonon, it is denoted as $\gamma_t$. The coupling constants $\gamma_{l}$ and $\gamma_t$ are adjustable parameters in TTLS model, and they were assumed to have no specific relation. However, in 1988 Meissner and Berret\cite{Berret1988} measured 18 different kinds of glass materials below the temperature of 1K, including chemically pured materials (e.g., a-SiO$_2$), chemically mixed materials (e.g., BK7) and organic materials (e.g., PMMA). They find that the ratios between $\gamma_l$ and $\gamma_t$ are not arbitrary: they range from 1.44 to 1.84, and most of them are around 1.5$\sim$1.6. Such observation suggests that the ratios $\gamma_l/\gamma_t$ are quite universal, regardless of the chemical compounds and microscopic structures of amorphous materials. TTLS model cannot explain this universality, because the model is based on the coupling constants. Therefore, we believe that there must be a more general model to describe the universal properties of low-temperature glasses, including the universal ratio $\gamma_l/\gamma_t$. In the rest of this paper, we use ``Meissner-Berret Ratio" to stand for ``the ratio between $\gamma_l$ and $\gamma_t$".

Besides this problem, there are a number of other problems in TTLS model. First, while TTLS successfully explained several universal propeties of amorphous solids below 1K, there are more universalities which cannot be explained within TTLS model around the temperature of 10K\cite{Leggett2011}, e.g. the plateau of thermal conductivity. Second, the TTLS model itself contains too many adjustable parameters. Experimental results could be fitted by adjusting these parameters. Third, the model lacks the consideration that as the interaction with phonon strain field, TTLS must generate a mutual RKKY-type interaction\cite{Joffrin1976}. Taking this virtual-phonon exchange interaction into account may not only change current theoretical results, but also question the validity of TTLS model. 

The purpose of this paper is to focus on the universality of Meissner-Berret ratio (the universal $\gamma_l/\gamma_t$ for various kinds of glasses) by developing a theory of coupled generic blocks. We start by expanding non-elastic part of glass Hamiltonian in orders of intrinsic and external phonon strain fields $\bm{e}(\vec x)$ and $\bm{e}(\vec x, t)$ to calculate the resonance phonon energy absorption for longitudinal and transverse external phonon fields. Within TTLS model, the resonance energy absorption per unit time $\dot{ E}_{l,t}$ is proportional to the square of coupling constants $\gamma_{l,t}$: $\dot{ E}_{l}/\dot{ E}_{t}=\gamma_l^2/\gamma_t^2$. In our generic coupled block model, the resonance energy absorption is proportional to the imaginary part of non-elastic susceptibility $\dot{ E}_{l}/\dot{ E}_{t}={\rm Im\,}\chi_l(\omega)/{\rm Im\,}\chi_t(\omega)$. Therefore, if we want to prove the universal property of $\gamma_l/\gamma_t$, we are actually proving the universal property of ${\rm Im\,}\chi_l(\omega)/{\rm Im\,}\chi_t(\omega)$. We combine a set of single blocks together to form a super block of glass. We allow the virtual phonons to exchange with each other to set up the RKKY-type many-body interaction between different single blocks. By putting in virtual phonon exchange interaction, we set up the renormalization recursion relation of resonance phonon energy absorption between single block and super block glasses. We repeat the real space renormalization procedure from starting microscopic length scale to obtain the resonance energy absorption at experimental length scales, and try to prove the universal ratio of ${\rm Im\,}\chi_l(\omega)/{\rm Im\,}\chi_t(\omega)$. Since the RKKY-type many-body interaction is independent of materials' microscopic structure and chemical compound, we believe that the ratio ${\rm Im\,}\chi_l(\omega)/{\rm Im\,}\chi_t(\omega)$ at experimental length scale will be able to explain the material-independent Meissner-Berret ratio. 

The organization of this paper is as follows: in section 2 we set up the main goal of this paper, the problem of universal Meissner-Berret ratio in TTLS model. We expand the general glass Hamiltonian in orders of intrinsic and external phonon strain fields, and introduce the most important concepts of this paper, namely the non-elastic stress tensor and non-elastic susceptibility. In section 3 we set up the renormalization recursion relation for the resonance phonon energy absorption between single block and super block glasses. In section 4 we repeat the renormalization process from microscopic length scale to experimental length scale and carry out the Meissner-Berret ratio, ${\rm Im\,}\chi_l(\omega)/{\rm Im\,}\chi_t(\omega)$. We prove at experimental length scale, the Meissner-Berret ratio ${\rm Im\,}\chi_l(\omega)/{\rm Im\,}\chi_t(\omega)$ is proportional to the sound velocity ratio $c_l/c_t$. We also use the least square method to investigate the statistical significance between the theory and experiment for the data of 13 amorphous materials listed in section 4. In section 5 we give a detailed discussion on the resonance phonon energy absorption contribution from the electric dipole-dipole interaction for dielectric amorphous solids. The contribution to Meissner-Berret ratio from electric dipole-dipole interaction is renormalization irrelevant. In Appendix (A) we give a detailed correction on the coefficient of non-elastic stress-stress interaction $\Lambda_{ijkl}^{(ss')}$, which was first obtained by Joffrin and Levelut\cite{Joffrin1976}.

\section{The Set up of Meissner-Berret Ratio Problem}

\subsection{The Definitions Non-elastic Hamiltonian, Stress Tensor and Susceptibility}
Let us consider a block of amorphous material. Since our purpose is to discuss the universal property of ``Meissner-Berret ratio" in amorphous materials below the temperature of 10K, we want to begin our investigation from the famous tunneling-two-level-system model (TTLS model)\cite{Phillips1987}. In this model we assume that there are a group of TTLSs randomly embedded in the glass material, with the location of the $i$-th TTLS at $\vec x_i$. With the presence of external phonon strain fields $\bm{e}(\vec x, t)$, the effective Hamiltonian $\hat{H}^{\rm tot}$ of amorphous material in TTLS model is the summation of long wavelength phonon Hamiltonian $\hat{H}^{\rm el}$, a group of tunneling-two-level-systems Hamiltonian, the couplings between two-level-systems and phonon intrinsic strain fields $\bm{e}(\vec x)$, and the couplings between two-level-systems and phonon external strain fields $\bm{e}(\vec x, t)$:
\begin{eqnarray}\label{1}
\hat{H}^{\rm tot}(t) & = & \hat{H}^{\rm el}+\sum_i\frac{1}{2}\left(
\begin{array}{cc}
E_i & 0\\
0 & -E_i\\
\end{array}
\right)\nonumber \\
 & + & \frac{\gamma_{l,t}}{2}\sum_i \left(
\begin{array}{cc}
D_i & M_i\\
M_i & -D_i\\
\end{array}
\right)\cdot\bm{e}(\vec x_i)\nonumber \\
 & + & \frac{\gamma_{l,t}}{2}Ak\sum_i \left(
\begin{array}{cc}
D_i & M_i\\
M_i & -D_i\\
\end{array}
\right)e^{i\vec k\cdot \vec x_i-i\omega t}
\end{eqnarray}
where the first, second, third and fourth terms stand for the long wavelength phonon Hamiltonian (we will also call it ``purely elastic Hamiltonian, $\hat{H}^{\rm el}$"), the Hamiltonian of a group of two-level-systems, the couplings between two-level-systems and intrinsic phonon strain fields $\bm{e}(\vec x)$, and the couplings between two-level-systems and external phonon strain fields $\bm{e}(\vec x, t)$. The two-level-systems Hamiltonian are written in the representation of energy eigenvalue basis, with the energy splitting $E_i=\sqrt{\Delta_i^2+\Delta_{0i}^2}$; $D_i=\Delta_i/E_i$ and $M_i=\Delta_{0i}/E_i$ are thr diagonal and off-diagonal matrix elements of the coupling between two-level-system and phonon strain field at location $\vec x_i$, and by definition they are no greater than 1; $\bm{e}(\vec x_i)$ is the local intrinsic phonon strain field at the position of the $i$-th two-level-system; $Ak$ is the product of external phonon strain field amplitude $A$ and external phonon strain field wave number $k$; $\omega$ is the frequency of external phonon strain field; $\gamma_{l,t}$ are the coupling constants between two-level-systems and longitudinal/transverse phonon strain fields.

Because the external phonon strain fields are coupled to two-level-systems, if the energy splitting of a certain two-level-system, $E_i$ matches $\hbar\omega$, then this two-level-system can resonantly absorb external phonon energy. By using Fermi golden rule, the resonance phonon energy absorption per unit time is proportional to the square of coupling constants:
\begin{eqnarray}\label{1.1}
{\dot{ E}_{l,t}} & = & \frac{\pi}{2\hbar} A^2k^2M_i^2E_i\tanh\left(\frac{1}{2}\beta\hbar\omega\right)\delta(E_i-\hbar\omega)\gamma_{l,t}^2\nonumber \\
 & \Rightarrow & \frac{\dot{E}_l}{\dot{E}_t}=\frac{ \gamma_{l}^2}{\gamma_t^2}
\end{eqnarray}
where $\dot{E}_l$ stands for the resonance phonon energy absorption for longitudinal external phonon fields, while $\dot{E}_t$ stands for the transverse phonon energy absorption. We assume that the phonon strain $e=Ak$ is identical for longitudinal and transverse external phonon strain fields. In TTLS model, $\gamma_{l,t}$ are assumed to be independent of each other. In other words, they have no specific relation.

However, in 1988, Meissner and Berret\cite{Berret1988} measureed 18 different kinds of glass materials, including chemically pured materials (for example, amorphous SiO$_2$), chemically mixed materials (for example, BK7) and organic materials (for example, PMMA and PS). They find that the ratio between longitudinal and transverse coupling constants $\gamma_l/\gamma_t$ ranges from 1.44 to 1.84. Most of the ratios are around 1.5$\sim$1.6. In the rest of this paper, let us name the ratio between $\gamma_l$ and $\gamma_t$ as ``Meissner-Berret ratio". TTLS model cannot explain such universality, because the model is based on the coupling constants. Therefore, we would like to believe, that there must be a more general model to describe the universal properties of low-temperature glasses, including the universal Meissner-Berret ratio.

In this subsection we want to set up a multiple-level-system model from the generalization of 2-level-system model. At this stage of setting up our model, we have not applied any external phonon strain field yet. We will consider external phonon strain fields in subsection 2(C). We begin our discussion by considering a single block of glass with the length scale $L$ much greater than the atomic distance $a\sim 10\AA$. In the subsection 2(B), we will combine a group of such single blocks to form a ``super block". We will consider the RKKY-type interaction between these single blocks, which is generated by virtual phonon exchange process. For now, we do not consider RKKY-type interaction and focus on the Hamiltonian of single block glass only.

We further define intrinsic phonon strain field $e_{ij}(\vec x)$ at position $\vec x$: if $\vec u(\vec x)$ denotes the displacement relative to some arbitrary reference frame of the matter at point $\vec x$, then strain field is defined as follows
\begin{eqnarray}\label{2}
e_{ij}(\vec x)=\frac{1}{2}\left(\frac{\partial u_i(\vec x)}{\partial x_j}+\frac{\partial u_j(\vec x)}{\partial x_i}\right)
\end{eqnarray}
We write down our general glass Hamiltonian as $\hat{H}^{\rm tot}$. Let us separate out from the glass general Hamiltonian $\hat{H}^{\rm tot}$, the purely elastic contribution $\hat{H}^{\rm el}$. It can be represented by phonon creation-annihilation operators as follows:
\begin{eqnarray}\label{3}
\hat{H}^{\rm el}
 = \sum_{k\alpha}\hbar\omega_{k\alpha}\left(\hat{a}_{k\alpha}^{\dag}\hat{a}_{k\alpha}+\frac{1}{2}\right)
\end{eqnarray}
where $\alpha=l,t$ represents phonon polarization, i.e., longitudinal and transverse phonons.

Subtracting the purely elastic part of Hamiltonian $\hat{H}^{\rm el}$, we name the left-over glass Hamiltonian $(\hat{H}^{\rm tot}-\hat{H}^{\rm el})$ as ``the non-elastic part of glass Hamiltonian, $\hat{H}^{\rm non}$". We expand the left-over Hamiltonian $\hat{H}^{\rm non}$ up to the first order expansion of long wavelength intrinsic phonon strain field. We name the coefficient of the first order expansion to be ``non-elastic stress tensor $\hat{T}_{ij}^{\rm non}(\vec x)$", defined as follows:
\begin{eqnarray}\label{4}
 & {} & \hat{H}^{\rm non} = \hat{H}^{\rm non}_0+\int d^3x\sum_{ij}e_{ij}(\vec x)\hat{T}^{\rm non}_{ij}(\vec x)+\mathcal{O}(e_{ij}^2)\nonumber \\
 & {} & \hat{T}_{ij}^{\rm non}(\vec x) = \frac{\delta \hat{H}^{\rm non}}{\delta e_{ij}(\vec x)}
\end{eqnarray}
Now let us compare Eq.(\ref{4}) with TTLS model Hamiltonian: the zeroth order expansion of non-elastic Hamiltonian $\hat{H}^{\rm non}$ with respect to strain field $e_{ij}$, $\hat{H}^{\rm non}_0$, is the generalization from two-level-system Hamiltonian to multiple-level-system Hamiltonian; non-elastic stress tensor $\hat{T}_{ij}^{\rm non}$ is the multiple-level generalization of the $2\times 2$ matrix which couples to phonon strain field in TTLS model. In the rest of this paper, we denote $\hat{H}^{\rm non}_0$ to be the non-elastic Hamiltonian excluding the coupling between intrinsic phonon strain $e_{ij}(\vec x)$ and non-elastic stress tensor $\hat{T}_{ij}^{\rm non}(\vec x)$. We denote $\hat{H}^{\rm non}$ to be the non-elastic Hamiltonian including the stress tensor $\hat{T}_{ij}^{\rm non}$--intrinsic phonon strain field $e_{ij}(\vec x)$ couplings (see the second equation of Eq.(\ref{4})).

Let us denote $|m\rangle$ and $E_m$ to be the $m$-th eigenstate and eigenvalue of the non-elastic Hamiltonian $\hat{H}_0^{\rm non}$. Such a set of eigenbasis $|m\rangle$ is a generic multiple-level-system. Now we can define the most important quantity of this paper, namely the non-elastic stress-stress susceptibility (i.e., linear response function). Let us apply an external infinitesimal testing strain, $e_{ij}(\vec x, t)=e_{ij}(\vec k)e^{i(\vec k\cdot \vec x-\omega_k t)}$. The non-elastic Hamiltonian of amorphous material, $\hat{H}^{\rm non}$, will provide a stress response $\langle \hat{T}_{ij}^{\rm non}\rangle(\vec x, t)=\langle \hat{T}_{ij}^{\rm non}\rangle e^{i(\vec k\cdot \vec x-\omega_kt)}$. Then we are ready to define the non-elastic stress-stress susceptibility (complex response function\cite{Anderson1986}) $\chi_{ijkl}^{\rm non}(\vec x, \vec x'; t, t')$
\begin{eqnarray}\label{5}
\chi_{ijkl}^{\rm non}(\vec x, \vec x'; \omega, \omega') & = & \int dtdt' \,e^{i\omega t+i\omega't'}\chi_{ijkl}^{\rm non}(\vec x, \vec x'; t, t') \nonumber \\
\chi_{ijkl}^{\rm non}(\vec x, \vec x'; t, t') & = & \frac{\delta  \langle\hat{T}^{\rm non}_{ij}\rangle(\vec x, t)}{\delta e_{kl}(\vec x',t')}
\end{eqnarray}
In the rest of this paper we will always use $\hat{H}$, $\hat{H}_0$, $\chi_{ijkl}$ and $\hat{T}_{ij}$ to represent non-elastic Hamiltonians $\hat{H}^{\rm non}$, $\hat{H}_0^{\rm non}$, susceptibility $\chi_{ijkl}^{\rm non}$ and stress tensor $\hat{T}^{\rm non}_{ij}$ respectively, while we use $\hat{H}^{\rm el}$, $\chi_{ijkl}^{\rm el}$ and $\hat{T}_{ij}^{\rm el}$ to represent the elastic Hamiltonian, susceptibility and stress tensor, respectively. In Eq.(\ref{5}) the stress response of non-elastic Hamiltonian, $\langle \hat{T}_{ij}\rangle(\vec x, t)$, is defined as follows: (please note from now on we use $\hat{T}_{ij}$ to stand for $\hat{T}_{ij}^{\rm non}$)
\begin{eqnarray}\label{6}
\langle \hat{T}_{ij}\rangle(\vec x, t) & = & \frac{\delta F(t)}{\delta e_{ij}(\vec x, t)}=-\frac{1}{\beta}\frac{\delta }{\delta e_{ij}(\vec x, t)}\ln \mathcal{Z}(t)\nonumber \\
 & = & \sum_m\frac{e^{-\beta E_m(t)}}{\mathcal{Z}(t)}\langle m_I,t|\hat{T}_{ij,(I)}(\vec x, t)|m_I,t\rangle\quad
\end{eqnarray}
where $\mathcal{Z}(t)=\sum_ne^{-\beta E(t)}$ is the time-dependent partition function of non-elastic Hamiltonian. With the presence of external testing strain field $e_{ij}(\vec x, t)$, the amorphous material receives a time-dependent perturbation $\int d^3x\,\sum_{ij}e_{ij}(\vec x, t)\hat{T}_{ij}(\vec x)$. In the representation in which $\hat{H}_0$ is diagonal, the perturbation has both of diagonal and off-diagonal matrix elements. The diagonal matrix elements of external perturabtion shift the energy eigenvalues: $E_n(t)= E_n+\int d^3x\,e_{ij}(\vec x, t)\langle n|\hat{T}_{ij}(\vec x) |n\rangle$, resulting in the shifts of probability function and partition function: $e^{-\beta E_n(t)}/\mathcal{Z}(t)$, $\mathcal{Z}(t)=\sum_ne^{-\beta E_n(t)}$. The off-diagonal matrix elements change the eigenstate wavefunctions: $|m_I, t\rangle=\mathcal{T}e^{\frac{1}{i\hbar}\int^t d^3x \, e_{ij}(\vec x, t')\hat{T}_{ij,(I)}(\vec x, t')dt'}|m\rangle$, where $|m_I, t\rangle$ is the wavefunction in the interaction picture, and $\hat{T}_{ij,(I)}(\vec x, t)$ is the stress tensor operator in the interaction picture: $\hat{T}_{ij,(I)}(\vec x, t)=e^{i\hat{H}_0t/\hbar}\hat{T}_{ij}(\vec x)e^{-i\hat{H}_0t/\hbar}$.

We further define the space-averaged non-elastic susceptibility for a single block of glass with the volume $V=L^3$, which will be very useful in later discussions,
\begin{eqnarray}\label{7.1}
& {} & \chi_{ijkl}(\omega) = \frac{1}{V}\int_{V} d^3xd^3x'\,\chi_{ijkl}(\vec x, \vec x'; \omega)
\end{eqnarray}
It is very useful to apply the assumption in the rest of this paper, that our amorphous material is invariant under real space SO(3) rotational group. Therefore, the non-elastic susceptibility obeys the generic form of an arbitrary 4-indice isotropic quantity: $\chi_{ijkl}=(\chi_l-2\chi_t)\delta_{ij}\delta_{kl}+\chi_t(\delta_{ik}\delta_{jl}+\delta_{il}\delta_{jk})$, where $\chi_l$ is the non-elastic part of glass compression modulus and $\chi_t$ is the non-elastic shear modulus.

According to the definitions Eqs.(\ref{5}, \ref{6}, \ref{7.1}), the space-averaged imaginary part of non-elastic susceptibility is given as follows, 
\begin{eqnarray}\label{11}
 & {} & {\rm Im}\,{\chi}_{ijkl}(T, \omega)=\sum_m\frac{e^{-\beta E_m}}{\mathcal{Z}}{\rm Im}\,{\chi}_{ijkl}^{(m)}(\omega)\nonumber \\
 & {} & {\rm Im}\,{\chi}_{ijkl}^{(m)}(\omega)=\frac{\pi}{L^3}\int d^3xd^3x'\sum_n\langle m|\hat{T}_{ij}(\vec x)|n\rangle\langle n|\hat{T}_{kl}(\vec x')|m\rangle\nonumber \\
 & {} &  \quad\quad\quad\quad\quad\quad \left[-\delta(E_n-E_m-\omega)+\delta(E_n-E_m+\omega)\right]
\end{eqnarray}
Where $\mathcal{Z}=\sum_ne^{-\beta E_n}$ is the  partition function, and we set $\hbar=1$. Please note that according to the definition of non-elastic susceptibility in Eq.(\ref{5}), the imaginary part of non-elastic susceptibility is negative for positive $\omega$. Because for arbitrary quantum number $n$ we always have $E_n\ge E_0$, the definition of ${\rm Im}\,\chi_{ijkl}^{(m)}(\omega)$ in Eq.(\ref{11}) is only valid when $E_m\ge \omega\ge -E_m$; when $E_m<\omega$ or $-E_m>\omega$, in Eq.(\ref{11}) one of the delta-functions will vanish. Therefore when $E_m<\omega$ or $-E_m>\omega$, the imaginary part of non-elastic susceptibility is simplified as follows, 
\begin{eqnarray}\label{11.1}
{\rm Im}\,{\chi}_{ijkl}^{(m)}(\omega) & = & \frac{\pi}{L^3}\int d^3xd^3x'\sum_n\langle m|\hat{T}_{ij}(\vec x)|n\rangle\langle n|\hat{T}_{kl}(\vec x')|m\rangle\nonumber \\
 & {} &   \times\delta(E_n-E_m+\omega)\quad {\rm if}\quad \omega<-E_m\nonumber \\
{\rm Im}\,{\chi}_{ijkl}^{(m)}(\omega) & = & \frac{\pi}{L^3}\int d^3xd^3x'\sum_n\langle m|\hat{T}_{ij}(\vec x)|n\rangle\langle n|\hat{T}_{kl}(\vec x')|m\rangle\nonumber \\
 & {} &   \times \left[-\delta(E_n-E_m-\omega)\right]\quad {\rm if}\quad \omega>E_m
\end{eqnarray}
Eq.(\ref{11.1}) is the supplemental definition of the imaginary part of non-elastic susceptibility. It is convenient to rewrite the imaginary part of non-elastic susceptibility in Eq.(\ref{11}) into the ``imaginary part of reduced non-elastic susceptibility ${\rm Im}\,\tilde{\chi}_{ijkl}$" as follows, for future use:
\begin{eqnarray}
 & {} & {\rm Im}\,{\chi}_{ijkl}(T, \omega)=\left(1-e^{-\beta \hbar\omega}\right){\rm Im}\,\tilde{\chi}_{ijkl}(T, \omega)\nonumber \\
 & {} & {\rm Im}\,\tilde{\chi}_{ijkl}(T, \omega) = \sum_m\frac{e^{-\beta E_m}}{\mathcal{Z}}{\rm Im}\,\tilde{\chi}_{ijkl}^{(m)}(\omega)\nonumber \\
 & {} & {\rm Im}\,\tilde{\chi}_{ijkl}^{(m)}(\omega) = \frac{\pi}{L^3}\int d^3xd^3x'\sum_n\langle m|\hat{T}_{ij}(\vec x)|n\rangle\langle n|\hat{T}_{kl}(\vec x')|m\rangle\nonumber \\
 & {} &   \qquad\qquad \quad\quad\times [-\delta(E_n-E_m-\omega)]
\end{eqnarray}
Please note, that by definition ${\rm Im\,}\tilde{\chi}_{ijkl}(T, \omega)$ is also a negative quantity for $\omega>0$. Again, for an arbitrary isotropic system, the imaginary part of reduced non-elastic susceptibility must satisfy the genetic isotropic form as well,  
\begin{eqnarray}\label{12}
{\rm Im}\,\tilde{\chi}_{ijkl}(T, \omega) & = & (\,{\rm Im}\,\tilde{\chi}_l(T, \omega)-2\,{\rm Im}\,\tilde{\chi}_t(T, \omega))\delta_{ij}\delta_{kl}\nonumber \\
 & {} & +\,{\rm Im}\,\tilde{\chi}_t(T, \omega)(\delta_{ik}\delta_{jl}+\delta_{il}\delta_{jk})
\end{eqnarray}
The newly-defined quantities ${\rm Im}\,\tilde{\chi}_{l,t}(T, \omega)$ are negative as well. Please note that we use ${\rm Im}\,\tilde{\chi}_{l,t}(T, \omega)$ to stand for the imaginary part of reduced non-elastic compression/shear moduli. The real part of reduced non-elastic susceptibility ${\rm Re}\,\tilde{\chi}_{l,t}(T, \omega)$ can be obtained by Kramers-Kronig relation from the imaginary part of it.

\subsection{Virtual Phonon Exchange Interactions}
Within single-block considerations, non-elastic stress tensor $\hat{T}_{ij}(\vec x)$ and non-elastic part of glass Hamiltonian $\hat{H}_0$ are simply generalizations from 2-level-system to multiple-level-system. There is not much difference between the multiple-level-system model and TTLS model. However, if we combine a set of $N_0^3$ single blocks together to form a ``super block", the interactions between single blocks must be taken into account. Since the non-elastic stress tensors are coupled to intrinsic phonon strain field, if we allow virtual phonons to exchange with each other, it will generate a RKKY-type many-body interaction between single blocks. This RKKY-type interaction is the product of stress tensors at different locations:
\begin{eqnarray}\label{13}
\hat{V}=\int d^3xd^3x'\sum_{ijkl}\Lambda_{ijkl}(\vec x-\vec x')\hat{T}_{ij}(\vec x)\hat{T}_{kl}(\vec x')
\end{eqnarray}
where the coefficient $\Lambda_{ijkl}(\vec x-\vec x')$ was first derived by Joffrin and Levelut\cite{Joffrin1976}. We give a further correction to it in Appendix (A).
\begin{eqnarray}\label{14}
{\Lambda}_{ijkl}(\vec x-\vec x') = -\frac{\tilde{\Lambda}_{ijkl}(\vec n)}{8\pi\rho c_t^2|\vec x-\vec x'|^3}
\end{eqnarray}
\begin{eqnarray}\label{15}
\tilde{\Lambda}_{ijkl} & = & \frac{1}{4}\bigg\{(\delta_{jl}-3n_jn_l)\delta_{ik}+(\delta_{jk}-3n_jn_k)\delta_{il}\nonumber \\
 & {} & +(\delta_{ik}-3n_in_k)\delta_{jl}
+(\delta_{il}-3n_in_l)\delta_{jk}\bigg\}\nonumber \\
 & {} & +\frac{1}{2}\left(1-\frac{c_t^2}{c_l^2}\right)\bigg\{-(\delta_{ij}\delta_{kl}+\delta_{ik}\delta_{jl}+\delta_{jk}\delta_{il})\nonumber \\
 & {} & +3(n_in_j\delta_{kl}+n_in_k\delta_{jl}+n_in_l\delta_{jk}\nonumber \\
 & {} & +n_jn_k\delta_{il}+n_jn_l\delta_{ik}+n_kn_l\delta_{ij})-15n_in_jn_kn_l\bigg\}\nonumber \\
\end{eqnarray}
where $\vec n$ is the unit vector of $\vec x-\vec x'$, and $i,j,k,l$ runs over $1,2,3$ cartesian coordinates. In the rest of this paper we will call Eq.(\ref{13}) non-elastic stress-stress interaction. In the rest of this paper we always use the approximation to replace $\vec x-\vec x'$ by $\vec x_{s}-\vec x_{s'}$ for the pair of the $s$-th and $s'$-th blocks, when $\vec x_{s}$ denotes the center of the $s$-th block, and that $\int_{V^{(s)}}\hat{T}_{ij}(\vec x)d^3x=\hat{T}_{ij}^{(s)}$ is the uniform stress tensor of the $s$-th block. Also, from now on we use $e_{ij}^{(s)}$ to denote the intrinsic phonon strain field $e_{ij}(\vec x)$ located at the $s$-th block. By combining $N_0\times N_0\times N_0$ identical $L\times L\times L$ single blocks to form a $N_0L\times N_0L\times N_0L$ super block, the Hamiltonian without the presence of external phonon strain fields is written as
\begin{eqnarray}\label{16}
\hat{H}^{\rm super}=\sum_{s=1}^{N_0^3}\hat{H}_0^{(s)}+\sum_{s\neq s'}^{N_0^3}\sum_{ijkl}\Lambda_{ijkl}^{(ss')}\hat{T}_{ij}^{(s)}\hat{T}_{kl}^{(s')}
\end{eqnarray}
From now on we apply the most important assumption, that these space-averaged stress tensors $\hat{T}_{ij}^{(s)}$ are diagonal in spacial coordinates: ${\rm Im}\,\tilde{\chi}^{(ss')}_{ijkl}(T,\omega)=\frac{1}{L^3}\langle \hat{T}_{ij}^{(s)}\hat{T}_{kl}^{(s')}\rangle ={\rm Im}\,\tilde{\chi}_{ijkl}(T, \omega)\delta_{ss'}$, which means for the stress tensors at different locations, they lose spacial correlations.

\subsection{Glass Non-elastic Hamiltonian with the Presence of External Strain field $\bm{e}(\vec x, t)$}
Next, let us consider the glass Hamiltonian with the presence of external weak strain field $\bm{e}(\vec x, t)$ as a perturbation. Please note that we have already used $\bm{e}(\vec x)$ to denote the intrinsic strain field, in this section we use $\bm{e}(\vec x; t)$ to stand for the external strain field. We further use $e_{ij}^{(s)}(t)$ to represent the external phonon strain field at the $s$-th single block of glass. It seems that the Hamiltonian Eq.(16) simply adds a stress-strain coupling term $\sum_s\sum_{ij}e_{ij}^{(s)}(t)\hat{T}_{ij}^{(s)}$. However, more questions arise with the appearance of external phonon strain field.

First of all these non-elastic stress tensors $\hat{T}_{ij}^{(s)}$ might be modified. A familiar example is that external phonon strain field can modify electric dipole moments by changing relative positions of positive-negative charge pairs (to the leading order of strain): $\Delta p_i(t)=\sum_{j}(\partial u_i(t)/\partial x_j)p_j$ where $i, j$ are cartesian coordinates, and $\vec u(\vec x, t)$ is the external phonon field. In principle we need to obtain the modification of stress tensors, $\Delta\hat{T}_{ij}^{(s)}(t)$ to the leading order in $e_{ij}^{(s)}(t)$ to calcuate the resonance phonon energy absorption correction. However, we only know the qualitative behavior of $\Delta\hat{T}_{ij}^{(s)}(t)$ in orders of external phonon strain field is $\Delta \hat{T}^{(s)}_{ij}(t)\sim e(t)\hat{T}_{ij}^{(s)}+\mathcal{O}(e^2)$. Therefore, we are only able to estimate the length scale dependence of the resonance phonon energy absorption correction from the contribution of $\Delta\hat{T}_{ij}^{(s)}(t)$. We will give the qualitative result of this correction in section 3, Eq.(\ref{25}). In section 4 we will show that this correction of resonance phonon energy absorption is renormalization irrelevant at experimental length scales  via scaling analysis.

There is a second problem arising from the presence of external phonon strain field: the relative positions of glass single blocks $\vec x_s-\vec x_{s'}$ can be changed, resulting in the modification of non-elastic stress-stress interaction coefficient $\Lambda_{ijkl}^{(ss')}(\bm{e})$. To the first order expansion in the external phonon strain field, the modification of $\Lambda_{ijkl}^{(ss')}$ is given as follows, 
\widetext
\begin{eqnarray}\label{20}
 & {} & \Delta\Lambda_{ijkl}^{(ss')} =  \left(\frac{x_{ss'}}{\Delta x_{ss'}}\Delta\tilde{\Lambda}^{(ss')}_{ijkl}-3\tilde{\Lambda}^{(ss')}_{ijkl}\cos\theta_{ss'}\right)\frac{\Delta x_{ss'}}{x_{ss'}^4}\nonumber \\
 & {} & \Delta\tilde{\Lambda}^{(ss')}_{ijkl} = \bigg\{\frac{3}{4}\bigg[2\big(n_jn_l\delta_{ik}+n_jn_k\delta_{il}+n_in_k\delta_{jl}+n_in_l\delta_{jk}\big)\cos\theta_{ss'}\nonumber \\
 & {} & -[(m_jn_l+m_ln_j)\delta_{ik}
+(m_jn_k+m_kn_j)\delta_{il}+(m_in_k+m_kn_i)\delta_{jl}+(m_in_l+m_ln_i)]\delta_{jk}
\bigg]\nonumber \\
 & {} & -3\alpha\cos\theta_{ss'}\bigg(n_kn_l\delta_{ij}+n_jn_l\delta_{ik}+n_kn_j\delta_{il}+n_in_l\delta_{jk}+n_in_k\delta_{jl}+n_in_j\delta_{kl}\bigg)\nonumber \\
 & {} & +\frac{3}{2}\alpha\bigg[m_i\left(n_l\delta_{jk}+n_k\delta_{jl}+n_j\delta_{kl}\right)+
m_j\left(n_l\delta_{ik}+n_k\delta_{il}+n_i\delta_{kl}\right)+m_k\left(n_l\delta_{ij}+n_i\delta_{jl}+n_j\delta_{il}\right)+
m_l\left(n_k\delta_{ij}+n_i\delta_{jk}+n_j\delta_{ik}\right)\bigg]\nonumber \\
 & {} & -\frac{15}{2}\alpha\bigg(m_in_jn_kn_l+m_jn_in_kn_l
+m_kn_in_jn_l+m_ln_in_jn_k\bigg)+30\alpha n_in_jn_kn_l\cos\theta_{ss'}\bigg\}\frac{\Delta x_{ss'}}{x_{ss'}}
\end{eqnarray}
where $\alpha=1-c_t^2/c_l^2$, $x_{ss'}=|\vec x_s-\vec x_s'|$, $\Delta \vec x_s=\vec u(\vec x_s, t)$, $\Delta x_{ss'}=|\Delta \vec x_s-\Delta \vec x_s'|$, $\cos\theta_{ss'}=(\Delta \vec x_{ss'}\cdot \vec x_{ss'})/\Delta x_{ss'}x_{ss'}$ is the angle between $\Delta \vec x_{ss'}$ and $\vec x_{ss'}$, and $\vec m=\Delta \vec x_{ss'}/\Delta x_{ss'}$ is the unit vector of $\Delta \vec x_{ss'}$. Finally the glass super block Hamiltonian with the presence of external phonon strain field $\bm{e}(\vec x, t)$ is given by
\begin{eqnarray}\label{21}
\hat{H}^{\rm super}(\bm{e}) & = & \sum_{s=1}^{N_0^3}\left(\hat{H}^{(s)}+\sum_{ij}e_{ij}^{(s)}(t)\hat{T}_{ij}^{(s)}\right)
+
\sum_{s\neq s'}^{N_0^3}\sum_{ijkl}\bigg(\Lambda_{ijkl}^{(ss')}\hat{T}_{ij}^{(s)}\hat{T}_{kl}^{(s')}+\Delta \Lambda_{ijkl}^{(ss')}(t)\hat{T}_{ij}^{(s)}\hat{T}_{kl}^{(s')}
+2\Lambda_{ijkl}^{(ss')}\Delta \hat{T}_{ij}^{(s)}(t)\hat{T}_{kl}^{(s')}\bigg)\nonumber \\
\end{eqnarray}

\endwidetext
\section{Resonance Phonon Energy Absorption of Super Block Glass}
According to the previous discussions, in TTLS model the resonance phonon energy absorption per unit time is proportional to the coupling constant squared: $\dot{ E}_{l,t}\propto \gamma_{l,t}^2$. In this section we want to use our generic coupled block model (i.e., the Hamiltonian in Eq.(\ref{21})) to calculate the same quantity. 

First of all, let us consider the $s$-th single-block glass with the dimension $L\times L\times L$, with the unperturbed non-elastic Hamiltonian $\hat{H}_0^{(s)}$ and time-dependent perturbation $\sum_{ij}e_{ij}^{(s)}(t)\hat{T}^{(s)}_{ij}$, so the total Hamiltonian of single block glass is $\hat{H}^{(s)}(t)=\hat{H}_0^{(s)}+\sum_{ij}e_{ij}^{(s)}(t)\hat{T}^{(s)}_{ij}$. We denote $|n^{(s)}\rangle$ and $E_n^{(s)}$ to be the $n^{(s)}$-th eigenstate and eigenvalue of unperturbed Hamiltonian $\hat{H}^{(s)}_0$. Thus the single-block resonance phonon energy absorption rate is $\dot{ E}_{l,t}^{\rm single}(L)=\frac{\partial}{\partial t}\sum_{n}\frac{e^{-\beta E_n^{(s)}}}{\mathcal{Z}^{(s)}}\left(\langle n^{(s)}_I, t|\hat{H}^{(s)}_I(t)|n^{(s)}_I,t\rangle-\langle n^{(s)}|\hat{H}^{(s)}_0|n^{(s)}\rangle\right)$, where $|n^{(s)}_I,t\rangle =\mathcal{T}e^{-\frac{i}{\hbar}\int_{-\infty}^t \sum_{ij}e^{(s)}_{ij}(t)\hat{T}^{(s)}_{ij(I)}(t') dt'}|n^{(s)}\rangle$ is the single block interaction picture wavefunction, and $\hat{H}^{(s)}_I(t)$ and $\hat{T}^{(s)}_{ij(I)}(t')$ are single block interaction picture operators. For an arbitrary operator $\hat{A}(t)$, the single block interaction picture operator is $\hat{A}_I(t)=e^{i\hat{H}^{(s)}_0t/\hbar}\hat{A}(t)e^{-i\hat{H}^{(s)}_0t/\hbar}$. The resonance phonon energy absorption per unit time of single-block glass is therefore given by 
\begin{eqnarray}\label{22}
\dot{ E}_{l,t}^{\rm single}(L) & = & -
2L^3A^2k^2\omega \left(1-e^{-\beta\hbar\omega}\right)\,{\rm Im}\,\tilde{\chi}_{l,t}(T,\omega)\nonumber \\
\end{eqnarray}
where $Ak$ is the product between external phonon strain field amplitude and phonon field wave number; $\omega$ is the frequency of external phonon field.  In Eq.(\ref{22}), according to the negativity of ${\rm Im}\,\tilde{\chi}_{l,t}(T,\omega)$, the single block energy absorption rate is positive. According to the argument by D. C. Vural and A. J. Leggett\cite{Leggett2011}, and the experiment by R. O. Pohl, X. Liu and E. Thompson\cite{Pohl2002}, we assume that below the temperature of 10K and within a certain range of frequency $\omega<\omega_c$, the imaginary part of reduced non-elastic susceptibilities are independent of frequency $\,{\rm Im}\,\tilde{\chi}_{l,t}(T, \omega)\approx \,{\rm Im}\,\tilde{\chi}_{l,t}(T)$. In section 4, Eq.(\ref{29}) we will discuss the order of magnitude of $\omega_c$ in details. Therefore, in our model, within single block considerations, the resonance phonon energy absorption is proportional to the imaginary part of reduced non-elastic susceptibility: $\dot{ E}_{l}^{\rm single}/\dot{ E}_{t}^{\rm single}=\,{\rm Im}\,\tilde{\chi}_{l}(T)/\,{\rm Im}\,\tilde{\chi}_t(T)$. If we want to prove the universal Meissner-Berret ratio, we are actually required to prove the universal property of ${\rm Im\,}\tilde{\chi}_l(T)/{\rm Im\,}\tilde{\chi}_t(T)$ in our model.

Let us combine $N_0^3$ glass non-interacting single blocks together to form a super block glass with the dimension $N_0L\times N_0L\times N_0L$. Such a group of non-interacting single blocks has the non-elastic part of glass Hamiltonian $\hat{H}_0=\sum_{s=1}^{N_0^3}\hat{H}_0^{(s)}$, eigenstates $|n\rangle = \prod_{s=1}^{N_0^3}|n^{(s)}\rangle$ and eigenvalues $E_n=\sum_{s=1}^{N_0^3}E_n^{(s)}$, where $|n^{(s)}\rangle$ and $E_n^{(s)}$ stand for the $n^{(s)}$-th eigenstate and eigenvalue for the $s$-th single block Hamiltonian $\hat{H}_0^{(s)}$. The partition function of these non-interacting single blocks is $\mathcal{Z}=\prod_s\mathcal{Z}^{(s)}$. We combine them to form a super block, and turn on non-elastic stress-stress interaction $\hat{V}$. We assume $\hat{V}$ is relatively weak compared to $\sum_{s=1}^{N_0^3}\hat{H}_0^{(s)}$, so it can be treated as a perturbation. Let us denote $|n^{\rm sup}\rangle$ and $E_n^{\rm sup}$ to be the $n^{\rm sup}$-th eigenstate and eigenvalue of super block static Hamiltonian $\hat{H}_0+\hat{V}$ ( $\sum_s\hat{H}^{(s)}_0+\hat{V}$ ). Their relations with $|n\rangle$ and $E_n$ are given as follows, 
\begin{eqnarray}
|n^{\rm sup}\rangle & = & |n\rangle+\sum_{p\neq n}\frac{\langle p|\hat{V}|n\rangle}{E_n-E_p}|p\rangle+\mathcal{O}(V^2)\nonumber \\
E_n^{\rm sup} & = & E_n+\langle n|\hat{V}|n\rangle +\mathcal{O}(V^2)
\end{eqnarray}
Finally, we turn on the time-dependent perturbation induced by external phonon: $\hat{H}'(t)=\sum_s\sum_{ij}e_{ij}^{(s)}(t)\hat{T}_{ij}^{(s)}+\sum_{ss'}\sum_{ijkl}\left(\Delta \Lambda_{ijkl}^{(ss')}(t)\hat{T}_{ij}^{(s)}\hat{T}_{kl}^{(s')}+2\Lambda_{ijkl}^{(ss')}\Delta \hat{T}_{ij}^{(s)}(t)\hat{T}_{kl}^{(s')}\right)$, to calculate the resonance phonon energy absorption rate by super block glass: 
\begin{eqnarray}\label{23}
 & {} & \dot{ E}_{l,t}^{\rm super} (L)= \frac{\partial}{\partial t}\sum_n\frac{e^{-\beta E_n^{\rm sup}}}{\mathcal{Z}^{\rm sup}}\nonumber \\
 & {} &  \left(\langle n^{\rm sup}_I, t|\hat{H}_{0I}(t)+\hat{V}_I(t)|n^{\rm sup}_I,t\rangle-\langle n^{\rm sup}|\hat{H}_0+\hat{V}|n^{\rm sup}\rangle\right)\nonumber \\
\end{eqnarray}
where $\mathcal{Z}^{\rm sup}=\sum_{n}e^{-\beta E_n^{\rm sup}}$ is the partition function for super block static Hamiltonian $\hat{H}_0+\hat{V}$; $|n^{\rm sup}_I,t\rangle =\mathcal{T}e^{-\frac{i}{\hbar}\int_{-\infty}^t \hat{H}_I'(t')dt'}|n^{\rm sup}\rangle$ is the super block interaction picture wavefunction, where $\hat{H}'(t)=\sum_s\sum_{ij}e_{ij}^{(s)}(t)\hat{T}_{ij}^{(s)}$; $\hat{H}'_I(t)$, $\hat{V}_I(t)$ and $\hat{H}_{0I}(t)$ are super block interaction picture operators: for arbitrary operator $\hat{A}(t)$, the super block interaction picture opertor is $\hat{A}_I(t)=e^{i(\hat{H}_0+\hat{V})t/\hbar}\hat{A}(t)e^{-i(\hat{H}_0+\hat{V})t/\hbar}$. 

We expand Eq.(\ref{23}) in orders of external phonon strain field $\bm{e}(\vec x, t)$ up to the second order. The first order expansion vanises after time-averaging. To the second order of $\bm{e}(\vec x, t)$, there are four contributions in the resonance phonon energy absorption rate. Three of them come from the time-dependent perturbation $\hat{H}'(t)$, while the last one comes from the non-elastic stress-stress interaction's contribution. We first consider the contribution of resonance phonon energy absorption rate from the external perturbation $\hat{H}'(t)$. There are three terms: the first contribution is the summation of single block energy absorption rate, 
\begin{eqnarray}\label{X1}
\dot{ E}_{l,t}^{(1)}(L)=N_0^3\dot{ E}_{l,t}^{\rm single}(L)
\end{eqnarray}
The second contribution comes from the expectation value of the operator which is quadratic in $\sum_{ss'}\sum_{ijkl}\Delta \Lambda_{ijkl}^{(ss')}(t)\hat{T}_{ij}^{(s)}\hat{T}_{kl}^{(s')}$:

\begin{eqnarray}\label{24}
 & {} & \dot{ E}_{l}^{(2)}(L)=\left(1-e^{-\beta\hbar\omega}\right)\frac{A^2k^2N_0^3\ln N_0}{40\pi^3(\rho c_t^2)^2}\nonumber \\
 & {} &  \left[(55+176\alpha+688\alpha^2)+44(1+4\alpha+4\alpha^2)x(T, \omega)\right]\nonumber \\ 
 & {} &\omega \int \,{\rm Im}\,\tilde{\chi}_t(T, \Omega)\,{\rm Im}\,\tilde{\chi}_t(T, \omega-\Omega){d\Omega }\nonumber \\
\nonumber \\
 & {} & \dot{ E}_{t}^{(2)}(L)=\left(1-e^{-\beta\hbar\omega}\right)\frac{A^2k^2N_0^3\ln N_0}{40 \pi^3(\rho c_t^2)^2}\nonumber \\
 & {} &  \left[(35+112\alpha+656\alpha^2)+28(1+4\alpha+4\alpha^2)x(T, \omega)\right]\nonumber \\  
& {} & \omega\int \,{\rm Im}\,\tilde{\chi}_t(T, \Omega)\,{\rm Im}\,\tilde{\chi}_t(T, \omega-\Omega){d\Omega }
\end{eqnarray}
where we denote $\alpha=1-{c_t^2}/{c_l^2}$ and $x(T, \omega)=\frac{\,{\rm Im}\,\tilde{\chi}_l(T, \omega)}{\,{\rm Im}\,\tilde{\chi}_t(T, \omega)}-2$. 
The third contribution to resonance phonon energy absorption is the expectation value of the operator quadratic in $2 \Lambda_{ijkl}^{(ss')}\Delta\hat{T}_{ij}^{(s)}(t)\hat{T}_{kl}^{(s')}$. Although we do not quantitatively know the first order expansion of stress tensor operator $\Delta\hat{T}_{ij}^{(s)}$ in orders of external phonon strain field, we can still use the qualitative relation $\Delta\hat{T}_{ij}\sim e \hat{T}_{ij}+\mathcal{O}(e^2)$ to estimate the length scale dependence of the third contribution to phonon energy absorption rate as follows,
\begin{eqnarray}\label{25}
 & {} & \dot{ E}_{l,t}^{(3)}(L)\sim K_{l,t}\left(1-e^{-\beta\hbar\omega}\right)\frac{A^2k^2N_0^3\ln N_0}{\pi^3(\rho c_{t}^2)^2}\nonumber \\
 & {} &\omega \int \,{\rm Im}\,\tilde{\chi}_t(T, \Omega)\,{\rm Im}\,\tilde{\chi}_t(T, \omega-\Omega){d\Omega }
\end{eqnarray}
 where $K_{l,t}$ are the constants for external longitudinal and transverse phonon strain fields, and they are of order $\sim 1$. By comparing Eq.(\ref{24}) with Eq.(\ref{25}), we find that the length scale dependences of $\dot{E}_{l,t}^{(2)}$ and $\dot{E}_{l,t}^{(3)}$ are the same. In the next section we will prove that they are both renormalization irrelevant throughout the entire renormalization procedure. 

Finally we consider the contribution of non-elastic stress-stress interaction $\hat{V}$ to the resonance phonon energy absorption: up to the second order expansion in terms of external phonon strain field $e_{ij}(\vec x, t)$, the contribution of energy absorption rate from $\hat{V}$ is given as follows
\begin{eqnarray}\label{26}
\dot{V}_{l,t}(L) & = & \left(1-e^{-\beta\hbar\omega}\right) \frac{4N_0^3L^3A^2k^2\ln N_0}{\rho c_{t,l}^2}\nonumber \\
 & {} & \omega\,{\rm Im}\,\tilde{\chi}_{t,l}(T,\omega){\rm Re}\,\tilde{\chi}_{t,l}(T,\omega)
\end{eqnarray}

The total resonance phonon energy absorption rate by super block glass is given by the summation of the above four terms. Because  the ``super block" glass in the $n$-th step renormalization, is actually the ``single block" glass in the $n+1$-th step of renormalization, we have the important relation $\dot{E}_{l,t}^{\rm single}(N_0L)=\dot{E}_{l,t}^{\rm super}(L)$. At length scale $L$, the ratio of super block resonance phonon energy absorption, is therefore given by $\dot{E}_{l}^{\rm single}(N_0L)/\dot{E}_{t}^{\rm single}(N_0L)$, which means the Meissner-Berret ratio is actually the functional of length scale. 
Finally, as a conclusion of this section, we write down the real space renormalization equation of resonance phonon energy absorption as follows, 
\begin{eqnarray}\label{X}
 & {} & \dot{E}_{l,t}^{\rm super}(L) = \dot{E}_{l,t}^{(1)}(L)+\dot{E}_{l,t}^{(2)}(L) +\dot{E}_{l,t}^{(3)}(L)+\dot{V}_{l,t}(L)\nonumber \\
 & {} & \dot{E}_{l,t}^{\rm single}(N_0L)=\dot{E}_{l,t}^{\rm super}(L)
\end{eqnarray}

\section{Real Space Renormalization of the Resonance Energy Absorption}
In this section we want to obtain the resonance phonon energy absorption rate and Meissner-Berret ratio at experimental length scales by repeating the real space renormalization procedure. Because the super block is $N_0^3$ times the volume of single block, repeating such renormalization process from microscopic length scale $L_1$ will eventually carry out the resonance phonon energy absorption rate at experimental length scale $R$. According to the argument by D. C. Vural and A. J. Leggett\cite{Leggett2011}, the suggested starting microscopic length scale of renormalization procedure is, for example, $L_1\sim 50\AA$. Since the final result only logarithmically depends on this choice, it will not be sensitive. In the $n$-th step of renormalization, we combine $N_0^3$ identical blocks with the side $L_n$ to form a $n$-th step super block with the side $N_0L_n$. Therefore, in the next step the single block length scale is $L_{n+1}=N_0L_n$.

We put in an external weak phonon strain field $\bm{e}(\vec x, t)$. The single and super block energy absorption rates in the $n$-th step renormalization are denoted as $\dot{E}_{l,t}^{\rm single}(L_n)$ and $\dot{E}_{l,t}^{\rm super}(L_n)$. According to the renormalization relation in Eq.(\ref{X}), we are able to set up the following renormalization relation between the $n$-th step and $n+1$-th step single block energy absorption:
\begin{eqnarray}\label{27}
 & {} & N_0^3\dot{E}_{l,t}^{\rm single}(L_n)+\dot{E}_{l,t}^{(2)} (L_n)+\dot{E}_{l,t}^{(3)}(L_n)+\dot{V}_{l,t}(L_n)\nonumber \\
 & {} & =\dot{E}_{l,t}^{\rm single}(L_{n+1})
\end{eqnarray}
It is very useful to define the ``energy absorption rate per volume" for future use: $\dot{\epsilon}_{l,t}(L_n)=L_n^{-3}\dot{ E}_{l,t}^{\rm single}(L_n)$, $\dot{\epsilon}_{l,t}^{(2, 3)}(L_n)=L_{n+1}^{-3}\dot{ E}_{l,t}^{(2, 3)}(L_n)$, $\dot{ v}_{l,t}(L_n)=L_{n+1}^{-3}\dot{ V}_{l,t}(L_n)$ and $\dot{\epsilon}_{l,t}(L_{n+1})=L_{n+1}^{-3}\dot{ E}_{l,t}^{\rm single}(L_{n+1})$. We repeat the renormalization procedure in Eq.(\ref{27}) for $\log_{N_0}(R/L_1)$ times from microscopic starting length scale $L_1\sim 50\AA$ to experimental length scale $R$. The energy absorption rate per volume at experimental length scale is given as follows 
\begin{eqnarray}\label{28}
 & {} & \dot{\epsilon}_{l,t}(R)\nonumber =\left(\dot{\epsilon}_{l,t}(L_1)+\dot{\epsilon}_{l,t}^{(2)}(L_1)+\dot{\epsilon}_{l,t}^{(3)}(L_1)\right)+\dot{v}_{l,t}\log_{N_0}\left(\frac{R}{L_1}\right)\nonumber \\
\end{eqnarray}

To study the properties of $\dot{\epsilon}_{l,t}(R)$ in Eq.(\ref{28}), we need to investigate all four terms on the r.h.s. of it. First, let us compare the length scale dependences of $\dot{ \epsilon}_{l,t}^{(2)}(L)$, $\dot{ \epsilon}_{l,t}^{(3)}(L)$ with $\dot{v}_{l,t}$:
\begin{eqnarray}\label{29}
\frac{\dot{ \epsilon}_{l,t}^{(2,3)}(L)}{\dot{ v}_{l,t}} & = & \frac{1}{\rho c_{l,t}^2L^3}\frac{\int\,{\rm Im}\, \tilde{\chi}_{l,t}(T,\Omega)\,{\rm Im}\,\tilde{\chi}_{l,t}(T,\omega-\Omega)d\Omega}{\,{\rm Im}\,\tilde{\chi}_{l,t}(T,\omega)\int\frac{\Omega\,{\rm Im}\,\tilde{\chi}_{l,t}(\Omega)d\Omega}{\Omega^2-\omega^2}}\nonumber \\
 & \sim & 
\frac{1}{\rho c_{l,t}^2L^3}\frac{\omega_c }{\int^{\omega_c}{\Omega}d\Omega/({\Omega^2-\omega^2})}\nonumber \\
 &\sim & 
\frac{1}{\rho c_{l,t}^2L^3}\frac{\omega_c }{\ln (\omega_c/\omega)}
\end{eqnarray}
If we require that there is a critical length scale $L_c$, beyond which $\dot{ \epsilon}_{l,t}^{(2,3)}(L_c)$ is smaller than $\dot{ v}_{l,t}$, then we need to estimate the order of magnitude for $L_c$. The upper limit of $L_c$ can be obtained by letting $\omega_c$ to take an extremely high value, $\omega_c\sim 10^{15}$rad/s which corresponds to $T=(\hbar\omega_c/k_B)\sim 10^4$K:
\begin{eqnarray}\label{23.5}
L_c< \left(\frac{1}{\rho c_{l,t}^2}\frac{\hbar\omega_c}{\ln\left({\omega_c}/{\omega}\right)}\right)^{\frac{1}{3}}\sim 1.7\AA< L_1=50\AA
\end{eqnarray}
The above analysis indicate that even if we choose an unreasonably huge cut-off frequency $\omega_c$, the upper limit of $L_c$ is still much smaller than the microscopic starting length scale of our generic coupled block model. Throughout the entire renormalization procedure $\dot{ \epsilon}_{l,t}^{(2,3)}(L)$ is always negligible compared to $\dot{ v}_{l,t}$. Therefore we are allowed to drop $\dot{ \epsilon}_{l,t}^{(2,3)}(L)$ in Eq.(\ref{28}).

Next let us compare the order of magnitude between $\dot{\epsilon}_{l,t}(L_1)$ and $\dot{v}_{l,t}\log_{N_0}\left({R}/{L_1}\right)$. The experimentally input ultrasonic phonon frequency is of the order $\omega\sim 10^7$rad/s, corresponding to the phonon wavelength $R\sim 10^{-3}$m. Hence  $\dot{v}_{l,t}\log_{N_0}\left({R}/{L_1}\right)$ is greater than $\dot{\epsilon}_{l,t}(L_1)$ for a factor of $\ln(R/L_1)\sim 20$. Therefore, we assume that the contribution of phonon energy absorption from single block Hamiltonian is much smaller than that from non-elastic stress-stress interaction: $\dot{\epsilon}_{l,t}(L_1)\ll \dot{v}_{l,t}\log_{N_0}\left({R}/{L_1}\right)$. At experimental length scales, the energy absorption rate per volume is dominated by the contribution of non-elastic stress-stress interaction, which is independent of the materials' microscopic nature. The ratio between the energy absorptions for longitudinal and transverse external phonon fields is given as follows:
\begin{eqnarray}\label{30}
\frac{\dot{\epsilon}_{l}(R)}{\dot{\epsilon}_{t}(R)}=\frac{\dot{v}_{l}}{\dot{ v}_{t}}
\end{eqnarray}
Note that on the r.h.s. of Eq.(\ref{30}), $\dot{v}_{l,t}$ is the product between the imaginary part and the real part of non-elastic susceptibilities (see Eq.(\ref{26})); on the other hand, the entire sample of glass block can be treated as a single gigantic block. The phonon energy absorption of this gigantic block is given by $\dot{ \epsilon}_{l,t}(R)=2A^2k^2\omega(1-e^{-\beta\hbar\omega})\,{\rm Im}\,\tilde{\chi}_{l,t}(T,\omega, R)$, with $\,{\rm Im}\,\tilde{\chi}_{l,t}(T,\omega, R)$ the reduced imaginary part of non-elastic susceptibility at experimental length scale $R$. Therefore, the l.h.s. of Eq.(\ref{30}) is given by $\frac{{\rm Im}\,\tilde{\chi}_{l}(T,\omega, R)}{{\rm Im}\,\tilde{\chi}_{t}(T,\omega, R)}$. Finally, we can self-consistently solve Eq.(\ref{30}) by requiring $\,{\rm Im}\,\tilde{\chi}_{l,t}(T,\omega, R)=\,{\rm Im}\,\tilde{\chi}_{l,t}(T,\omega)$. The only parameter which enters into this self-consistent equation is the ratio of sound velocity $c_l/c_t$, and it is an experimentally measurable quantity which cannot be adjusted arbitrarily.

The self-consistent solution is $\sqrt{\frac{\,{\rm Im}\,\tilde{\chi}_l(T, \omega)}{\,{\rm Im}\,\tilde{\chi}_t(T, \omega)}}=\frac{c_l}{c_t}$, which means in our generic coupled model, the theoretical prediction of Meissner-Berret ratio equals to the ratio of sound velocity. This result is in fairly good agreement with the 13 materials measured by Meissner and Berret\cite{Berret1988}. Other 5 materials in their paper, however, lack either the longitudinal or transverse sound velocity. We are not able to justify the correctness of our theory for those 5 materials. Therefore, we only list the 13 materials' data measured by Meissner and Berret in the following table. We use ``$ {\rm MB}_{\rm exp}=\gamma_l/\gamma_t $" to stand for the experimental Meissner-Berret ratio, while we use ``${\rm MB}_{\rm theo}=c_l/c_t $" to represent our theoretical prediction of Meissner-Berret ratio: \widetext{
\begin{displaymath}
\begin{array}{c|c|c|c|c|c|c|c|c|c|c|c|c}
 {\rm Material} & \gamma_l(\rm eV) & \gamma_t(\rm eV) & {\rm MB}_{\rm exp}=\gamma_l/\gamma_t &   c_l({\rm km/s}) & c_t({\rm km/s}) & {\rm MB}_{\rm theo}=c_l/c_t  & \frac{{\rm theo}-{\rm exp}}{\rm exp} \\
\hline
{\rm a\textrm{-}SiO_2} & 1.04 & 0.65 & 1.60  &  5.80 & 3.80 & 1.53 & -4.38\%\\ 
\hline
{\rm BK7}  & 0.96 & 0.65 & 1.48 &  6.20 & 3.80 & 1.63 & +10.1\% \\
\hline
{\rm As_2S_3 }  & 0.26 & 0.17 & 1.53 &  2.70 & 1.46 & 1.85 & +20.9\% \\
\hline
{\rm LaSF\textrm{-}7}  & 1.46 & 0.92 & 1.59 &  5.64 & 3.60 & 1.57 & -1.26\%\\
\hline
{\rm SF4}  & 0.72 & 0.48 & 1.50 &  3.78 & 2.24 & 1.69 & +12.7\%\\
\hline
{\rm SF59}  & 0.77 & 0.49 & 1.57 &  3.32 & 1.92 & 1.73 & +10.2\%\\
\hline
{\rm V52}  & 0.87 & 0.52 & 1.67 &   4.15 & 2.25 & 1.84 & +10.4\%\\
\hline
{\rm BALNA}  & 0.75 & 0.45 & 1.67 &  4.30 & 2.30 & 1.87 & +12.0\%\\
\hline
{\rm LAT}  & 1.13 & 0.65 & 1.74 &  4.78 & 2.80 & 1.71 & -1.72\%\\
\hline
{\rm a\textrm{-}Se}  & 0.25 & 0.14 & 1.79 &  2.00 & 1.05 & 1.90 & +6.14\%\\
\hline
{\rm Zn\textrm{-}Glass}  & 0.70 & 0.38 & 1.84 &  4.60 & 2.30 &  2.00 & +8.70\%\\
\hline
{\rm PMMA}  & 0.39 & 0.27 & 1.44 &  3.15 & 1.57 & 2.01 & +39.6\%\\
\hline
{\rm PS}  & 0.20 & 0.13 & 1.54 &  2.80 & 1.50 & 1.87 & +21.4\% \\
\hline
\end{array}
\end{displaymath}
\endwidetext
Among the above 13 materials, the theoretical predictions of As$_2$S$_3$, PMMA and PS deviate more than $20\%$ compared to their experimental data. We will give a further discussion regarding such great deviations in the next paragraph. For now let us investigate the statistical significance between the theoretical and experimental Meissner-Berret ratios. We use the least square method to find the best linear fitting between the theory and the experiment. For the above 13 materials including large theoretical deviations of As$_2$S$_3$, PMMA and PS, the least square fitting is $\left({\rm MB}\right)_{\rm theo}=1.102\left({\rm MB}\right)_{\rm exp}$, with the linear correlation coefficient $r=0.261$. This result indicates that the linear fitting is not a good description for the data of 13 materials; on the other hand, we use least square fitting for the 10 materials excluding As$_2$S$_3$, PMMA and PS. The least square fitting gives $\left({\rm MB}\right)_{\rm theo}=1.061\left({\rm MB}\right)_{\rm exp}$, with the linear correlation coefficient $r=0.745$. This result indicates that except for the large theoretical deviations of As$_2$S$_3$, PMMA and PS, the sound velocity ratio $c_l/c_t$ is a moderate theoretical description for the other 10 materials. We plot the theory versus experiment as follows, where $x$ and $y$-axis stand for experimental and theoretical Meissner-Berret ratios:
\begin{figure}[h]
\includegraphics[scale=0.426]{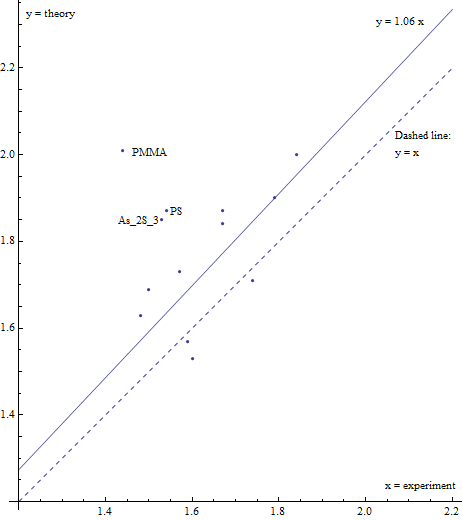}
\caption{The least square fitting for the Meissner-Berret ratios between the theory and the experiment. The least square fitting is $y=1.06x$ for the data of 10 materials excluding As$_2$S$_3$, PMMA and PS. The linear correlation coefficient is $r=0.745$. The dashed line is our original theoretical prediction of Meissner-Berret ratio ${\rm MB}_{\rm theo}=c_l/c_t$.}  
\end{figure}

Let us begin to discuss the large theoretical deviations of Meissner-Berret ratios for As$_2$S$_3$, PMMA and PS. Instead of measuring the resonance phonon energy absorption, the original experiment by Meissner and Berret\cite{Berret1988} was to measure the sound velocity shift as the function of temperature below 10K: $\Delta c_{l,t}/c_{l,t}=\mathcal{C}_{l,t}\ln (T/T_0)$. In their original paper, they used TTLS model to derive the coefficient $\mathcal{C}_{l,t}=\bar{P}\gamma_{l,t}^2/\rho c_{l,t}^2$ for longitudinal and transverse phonons theoretically, then they measured the TTLS parameter $\bar{P}$ to calculate the coupling constants $\gamma_{l,t}$. The definition of TTLS parameter $\bar{P}$ is given as follows\cite{Hunklinger1986}: in two-level-system, the diagonal matrix element $\Delta$ and the tunneling parameter $\lambda=\ln(\hbar\Omega/\Delta_0)$ are assumed to be independent of each other and to have a constant distribution $P(\Delta,\lambda)d\Delta d\lambda=\bar{P}d\Delta d\lambda$. Therefore the experimentally inferred Meissner-Berret ratios $\gamma_l/\gamma_t=\sqrt{\mathcal{C}_lc_l^2/\mathcal{C}_tc_t^2}$ are based on the assumptions of TTLS model: (1) whether these tunneling systems are strictly two-level-systems? (2) whether the parameters $\Delta$ and $\lambda$ are strictly independent of each other? (3) whether the parameters have a strict constant distribution? The experimentally inferred Meissner-Berret ratios are very sensitive to these assumptions.

Besides the problems we mentioned above, the validity of TTLS model itself might be questionable as well. To check the validity of TTLS model for As$_2$S$_3$, PMMA and PS, we search the experimental data of low-temperature heat capacity: As$_2$S$_3$ measured by R. B. Stephens\cite{Stephens1973}; PMMA and PS measured by R. B. Stephens, G. S. Cieloszyk and G. L. Salinger\cite{Stephens1972}; PMMA measured by R. C. Zeller and R. O. Pohl\cite{Zeller1971}. At low-temperatures below $1$K, the temperature dependences of their heat capacities largely deviate from the ``linear temperature dependence"  obtained by TTLS model, $C_v(T)=AT+BT^3$, where $A$ and $B$ are experimentally determined parameters. The huge deviations of heat capacities imply that TTLS model itself may not be a suitable description for As$_2$S$_3$, PMMA and PS below $1$K.  We suggest this might be one of the reasons to explain the large deviations of our theory.

In 2010, Ivan Y. Eremchev, Yury G. Vainer, Andrei V. Naumov and Lothar Kador\cite{Kador2010} used single-molecule (SM) spectroscopy to directly observe the existence of tunneling-two-level-systems in polymer glass with different internal structures and chemical compositions. Their observations demonstrate that TTLS model is not universal for low-temperature glasses: for high-molecular-weight polymer glasses, the SM spectroscopies present clear and stable two-level-system behaviors, while for those low-molecular-weight polymer glasses with the same chemical compound, the SM sepctroscopies are so fast and chaotic that SM spectra could hardly or not at all be recorded. They only observe irregular fluorescence flares. They observed SM spectroscopies for different kinds of glass materials, including: (1) high-molecular-weight materials (for example, Polyisobutylene with $M_w$=33800 g/mol, $M_w=$420000 g/mol). They present clear and stable SM sepctra. They perform jumps between a small number of spectral positions, but still, this number appears not to be a power of two in all cases; (2) low-molecular-weight materials (for example, Polyisobutylene with $M_w=390 $g/mol, 2500 g/mol, toluene with $M_w=92$ g/mol, Cumene with $M_w=120$g/mol, Propylene carbonate with $M_w=$102 g/mol, and oligome, a molecular complex that consists of a few monomer units). These materials present unstable and chaotic SM spectra which cannot be recorded clearly. The authors pointed out, that the polymer chains must contain more than hundreds of monomers so that the connection is strong enough to suppress the fast dynamics of glass molecules. Therefore, it is necessary to investigate the molecular weight of PMMA and PS samples used in the original experiment by Meissner and Berret. However, we couldn't retrieve the original molecular weight of PMMA and PS, because the samples were from G. Federle's thesis which was not published\cite{Federle1983}. 
\begin{figure}[h]
\includegraphics[scale=0.4]{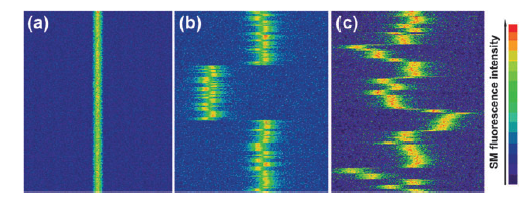}
\caption{(a)The typical behavior of crystalline solids\cite{Gorshelev2010}: single stable line; (b)single-molecule spectra\cite{Kador2010} for high-molecular weight glass Polyisobutylene with $M_w=420000$g/mol; (c) low-molecular-weight glass, frozen toluene, rich and complex SM spectra.}  
\end{figure}
\begin{figure}[h]
\includegraphics[scale=0.4]{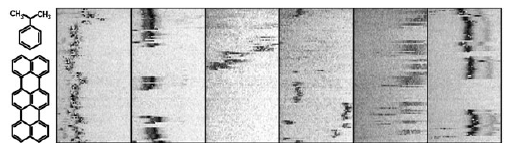}
\caption{Single-molecule spectra\cite{Kador2010} for low-molecular weight cumene. The SM spectra is unstable and hard to read.}  
\end{figure}
\begin{figure}[h]
\includegraphics[scale=0.4]{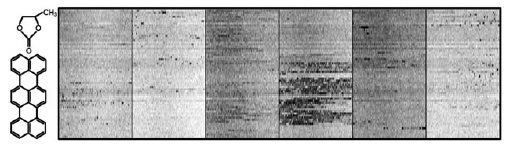}
\caption{Single-molecule spectra\cite{Kador2010} for low-molecular weight glass, propylene carbonate. The SM spectra is even worse.}  
\end{figure}

Besides the above analysis, we also suggest the possibility that As$_2$S$_3$, PMMA and PS may not possess the universal properties of low-temperature glass which could be observed in typical glass materials (e.g. a-SiO$_2$, (KCl)$_x$(KCN)$_{1-x}$, etc.), for example, the universal property of quadratic temperature dependence of thermal conductivity $\kappa\sim T^2$ below the temperature of $1$K\cite{Stephens1973}, and the universal plateau of thermal conductivity between 4K and 20K in glass materials\cite{Stephens1973}. We hope more measurements of Meissner-Berret ratios could be carried out to further investigate the correctness of our generic coupled block model and the TTLS model.

\section{The Modification of Meissner-Berret Ratio from Electric Dipole-Dipole Interactions}
Electric dipole moments interact with each other via a $1/r^{3}$ long range interaction which is similar with non-elastic stress-stress interactions. The external phonon waves (not electromagnetic waves) can change the relative positions $\vec x_s-\vec x_{s'}$ of different electric dipole moments at $\vec x_s$ and $\vec x_{s'}$; on the other hand, electric dipole moment is proportional to the separation of positive-negative charges: $\vec p=q\vec l$. Thus external phonons also modify electric dipole moments by changing the charge separation $\vec l\to \vec l+\Delta \vec l$. In conclusion, external phonons will change electric dipole-dipole interations, resulting in the change of phonon energy absorption by dielectric glass materials. 

However, as we will see at the end of this section, the influence of electric dipole-dipole interaction on phonon energy absorption is renormalization irrelevent, because of the following reason. From section 3, the external time-dependent perturbation $\hat{H}'(t)$ and non-elastic stress-stress interaction $\hat{V}$ have four contributions to the resonance phonon energy absorption rate $\dot{E}_{l,t}$: Eq.(\ref{X1}), Eqs.(\ref{24}), Eq.(\ref{25}) and Eq.(\ref{26}). Eqs.(\ref{24}) and Eq.(\ref{25}) are renormalization irrlevant, while Eq.(\ref{26}) is the only renormalization relevant term. Let us give a short review on Eqs.(\ref{24}, \ref{25}, \ref{26}): Eqs.(\ref{24}) is generated by the change of non-elastic stress-stress interaction coefficient $\Delta\Lambda_{ijkl}^{(ss')}(\bm{e})=\Lambda_{ijkl}^{(ss')}(\bm{e})-\Lambda_{ijkl}^{(ss')}$ due to the external strain field; Eq.(\ref{25}) is generated by the change of non-elastic stress tensor operator $\Delta\hat{T}_{ij}=\hat{T}_{ij}(\bm{e})-\hat{T}_{ij}$. These two terms generate phonon energy absorptions which are renormalization irrelevant; Eq.(\ref{26}) is generated by the change of wave function of single block glass Hamiltonian $\delta|n\rangle$ due to the coupling between stress tensor and external phonon strain field $e_{ij}\hat{T}_{ij}$. This term is renormalization relevant. When we consider the resonance phonon energy absorption contribution from the electric dipole-dipole interaction, we get the following contributions: (1) electric dipole-dipole interaction coefficient $\mu_{ij}^{(ss')}$ could be modified by external phonon field, thus we have the change of interaction coefficient $\Delta \mu_{ij}^{(ss')}=\mu_{ij}^{(ss')}(\bm{e})-\mu_{ij}^{(ss')}$. The contribution to phonon energy absorption from this term is renormalization irrelevant, which is similar with Eq.(\ref{24}); (2) electric dipole moments $\hat{p}_i$ could be changed by external phonon strain field, resulting in the change $\Delta\hat{p}_i=\hat{p}_i(\bm{e})-\hat{p}_i$. The phonon energy absorption contribution from this term is also renormalization irrelevant, similar with Eq.(\ref{25}); (3) similar with the stress-strain coupling $e_{ij}\hat{T}_{ij}$, electric field couples to electric dipole moment as well ( $-\vec E\cdot \vec {p}$ ). However, there is not such a term, that the elastic strain field couples to electric dipole moment. The resonance phonon energy absorption contribution from $e_{ij}\hat{T}_{ij}$ is the only renormalization relevant term. Such a kind of energy absorption does not exist when considering the coupling between external phonon strain field and electric dipole moments. 

Even though the contribution of phonon energy absorption from electric dipole-dipole interaction is renormalization irrlevant, we still give a quantitative analysis as follows. Let us first compare the order of magnitude between electric dipole-dipole interaction and non-elastic stress-stress interaction. We use $M$ to denote the off-diagonal matrix element of two-level-system and we use $n_0$ to denote the density of states for two-level-system. We also use $\mu$ to denote the electric dipole moment and use $n_e$ to denote the density of states for the two-level-system of electric dipole moment. Therefore, the magnitude of non-elastic susceptibility is ${n_0M^2}/{\rho c^2}$, while the magnitude of dielectric susceptibility is ${n_e\mu^2}/{\epsilon}$. Since the ``resonance phonon energy absorption rate" $\dot{E}_{l,t}$ is proportional to the imaginary part of non-elastic and dielectric susceptibilities, we need to compare ${n_e\mu^2}/{\epsilon}$ with ${n_0M^2}/{\rho c^2}$.

According to the data measured by S. Hunklinger and M. V. Schickfus\cite{Schickfus1981}, we discuss the ratio between ${n_e\mu^2}/{\epsilon}$ and ${n_0M^2}/{\rho c^2}$ for two dielectric materials below, BK7 and amorphous SiO$_2$.

For BK7, the parameters of TTLS model are of order $n_0M^2\sim 10^{8}$erg/cm$^3$; dielectric constant $\epsilon=3.7$; the electric dipole moment is of order $n_e\mu^2=6\times 10^{-3}$; mass density $\rho=2.51 $g/cm$^3$; sound velocity $c=6.5\times 10^5$cm/s. From these data we find the ratio between dielectric susceptibility and non-elastic susceptibility is 
$
({n_e\mu^2}/{\epsilon}:{n_0M^2}/{\rho c^2})\sim (1.62\times 10^{-3}: 0.94\times 10^{-4})
$, which means the strength of electric dipole-dipole interaction is one order of magnitude greater than that of non-elastic stress-stress interaction in BK7. 

For SiO$_2$, the parameters of TTLS model is $n_0M^2= 2.04\times 10^{8}$erg/cm$^3$; $\epsilon=3.81$; the parameters of electric dipole moment is $n_e\mu^2=1.46\times 10^{-4}$; $\rho=2.2 $g/cm$^3$; $c=5.8\times 10^5$cm/s; dielectric susceptibility versus non-elastic susceptibility is of order
$
({n_e\mu^2}/{\epsilon}:{n_0M^2}/{\rho c^2})\sim (3.83\times 10^{-5}: 2.76\times 10^{-4})
$, which means the strength of electric dipole-dipole interaction is one order of magnitude smaller than that of non-elastic stress-stress interaction in SiO$_2$.

The above qualitative analysis suggests that the electric dipole-dipole interaction in dielectric materials are roughly of the same order of magnitude as the non-elastic stress-stress interaction. However, after detailed calculations we will prove that the energy absorption contribution from electric dipole-dipole interaction is renormalization irrelevant, although the strength of electric dipole-dipole interaction is of the same order of magnitude. We use the approximation to replace $\vec x-\vec x'$ by $\vec x_{s}-\vec x_{s'}$ for the pair of the $s$-th and $s'$-th blocks, when $\vec x_{s}$ denotes the center of the $s$-th block, and $\int_{V^{(s)}}\hat{p}_{i}(\vec x)d^3x=\hat{p}_{i}^{(s)}$ is the uniform electric dipole moment for the $s$-th block. We also denote $\hat{H}_0^{(s)}$ to be the single block Hamiltonian for the $s$-th glass single block, and use $|n^{(s)}\rangle$ and $E_n^{(s)}$ to be the $n^{(s)}$-th eigenstate and eigenvalue of the single block Hamiltonian $\hat{H}_0^{(s)}$. By combining $N_0\times N_0\times N_0$ identical $L\times L\times L$ single blocks to form a $N_0L\times N_0L\times N_0L$ super block, the electric dipole-dipole interaction by super block glass is written as follows,
\begin{eqnarray}\label{31}
\hat{V}_{\rm dipole}=\sum_{s\neq s'}^{N_0^3}\sum_{i,j=1}^{3}\mu_{ij}^{(ss')}\hat{p}_{i}^{(s)}\hat{p}_{j}^{(s')}
\end{eqnarray}
where in the above Eq.(\ref{31}) we define the coefficient $\mu^{(ss')}_{ij}$ of electric dipole-dipole interaction as follows
\begin{eqnarray}\label{32}
\mu_{ij}^{(ss')}=\frac{(\delta_{ij}-3n_{i}n_{j})}{8\pi\epsilon |\vec x_s-\vec x_s'|^3}
\end{eqnarray}
in Eq.(\ref{31}, \ref{32}), the indiccecs $i, j$ run over $1,2,3$ cartesian coordinates and $\vec n$ is the unit vector of $\vec x_s-\vec x_{s'}$. The external phonon field $\vec u(\vec x, t)$ can modify (1)  the coefficient of electric dipole-dipole interaction $\mu_{ij}^{(ss')}$ by changing the relative positions of glass single blocks $\vec x_s-\vec x_{s'}$. We deonote it as $\Delta\mu_{ij}^{(ss')}$:
\begin{eqnarray}\label{33}
\Delta\mu_{ij}^{(ss')}=\frac{3\Delta x_{ss'}}{8\pi\epsilon x_{ss'}^4}\left[\left(5n_in_j-\delta_{ij}\right)\cos\theta_{ss'}-(n_jm_i+n_im_j)\right]\nonumber \\
\end{eqnarray}
where $x_{ss'}=|\vec x_s-\vec x_s'|$, $\Delta \vec x_s=\vec u(\vec x_s, t)$, $\Delta x_{ss'}=|\Delta \vec x_s-\Delta \vec x_s'|$, $\cos\theta_{ss'}=(\Delta \vec x_{ss'}\cdot \vec x_{ss'})/\Delta x_{ss'}x_{ss'}$ and $\vec m=\Delta \vec x_{ss'}/\Delta x_{ss'}$.
(2) The external phonon field can also change the electric dipole operators $\hat{p}^{(s)}$, because the positive and negative charges in a certin electric dipole are driven away from their original positions $\vec x_s\pm\frac{1}{2}\vec l_s$ to new positions $\vec x_s\pm\frac{1}{2}\vec l_s+\vec u(\vec x_s\pm\frac{1}{2}\vec l_s, t)$, leading to the change of electric dipole moment operators $\Delta\hat{p}^{(s)}$
\begin{eqnarray}\label{34}
\Delta \hat{p}_i(\vec x, t)=\sum_{k}\frac{\partial u_i(\vec x, t)}{\partial x_k}\hat{p}_k(\vec x)
\end{eqnarray}
Therefore with the presence of external phonon strain field $\bm{e}(\vec x, t)$, the electric dipole-dipole interaction is 
\begin{eqnarray}\label{35}
\hat{V}_{\rm dipole}(\bm{e}) & = & \sum_{s\neq s'}^{N_0^3}\sum_{i,j=1}^{3}\bigg(\mu_{ij}^{(ss')}\hat{p}_{i}^{(s)}\hat{p}_{j}^{(s')}
+\Delta\mu_{ij}^{(ss')}(t)\hat{p}_{i}^{(s)}\hat{p}_{j}^{(s')}\nonumber \\
 & {} & +2\mu_{ij}^{(ss')}\Delta\hat{p}_{i}^{(s)}(t)\hat{p}_{j}^{(s')}
\bigg)
\end{eqnarray}
Let us define the dielectric susceptibility $\chi_{ij}(T, \omega)$ for future use:
\begin{eqnarray}\label{36}
{\rm Im}{\chi}_{ij}(T, \omega) & = & \left(1-e^{-\beta\hbar\omega}\right)\,{\rm Im}\,{\tilde{\chi}}_{ij}(T, \omega)\nonumber \\
{\rm Im}\tilde{\chi}_{ij}(T, \omega) & = & \sum_m\frac{e^{-\beta E_m}}{\mathcal{Z}}\,{\rm Im}\,\tilde{\chi}_{ij}^{(m)}(\omega)\nonumber \\
{\rm Im}\tilde{\chi}_{ij}^{(m)}(\omega) & = & -\frac{\pi}{L^3}\sum_n\langle m|\hat{p}_i^{(s)}|n\rangle \langle n|\hat{p}_j^{(s)}|m\rangle\nonumber \\
 & {} & \delta(E_n-E_m-\omega)
\end{eqnarray}
where the imaginary part of dielectric susceptibility is by definition negative. Since the dielectric susceptibility must be invariant under SO(3) group, it takes the generic isotropic form $\,{\rm Im}\,\tilde{\chi}_{ij}(T,\omega )=\,{\rm Im}\,\tilde{\chi}(T, \omega)\delta_{ij}$.

To calculate the phonon energy absorption from the contribution of electric dipole-dipole interaction, we use the following two steps: (1) we combine $N_0^3$ non-interacting glass single blocks together to form a super block with the Hamiltonian $\hat{H}_0=\sum_s\hat{H}_0^{(s)}$. We denote $|n\rangle=\prod_{s}|n^{(s)}\rangle$ and $E_n=\sum_sE_{n}^{(s)}$ to be the  eigenstates and eigenvalues for $\hat{H}_0$; (2) we turn on the time-dependent perturbation generated by the external phonon strain field $\hat{H}'(t)=\sum_s\sum_{ij}e_{ij}^{(s)}(t)\hat{T}_{ij}^{(s)}+\sum_{ss'}\sum_{ijkl}\left(\Delta \Lambda_{ijkl}^{(ss')}(t)\hat{T}_{ij}^{(s)}\hat{T}_{kl}^{(s')}+2\Lambda_{ijkl}^{(ss')}\Delta \hat{T}_{ij}^{(s)}(t)\hat{T}_{kl}^{(s')}\right)+\sum_{ss'}\sum_{ij}\left(\Delta\mu_{ij}^{(ss')}(t)\hat{p}_{i}^{(s)}\hat{p}_{j}^{(s')}+2\mu_{ij}^{(ss')}\Delta\hat{p}_{i}^{(s)}(t)\hat{p}_{j}^{(s')}
\right)$; (3) we turn on the non-elastic stress-stress interaction and electric dipole-dipole interaction $\hat{V}+\hat{V}_{\rm dipole}$. The resonance phonon energy absorption is given by 
\widetext
\begin{eqnarray}
 & {} & \dot{ E}_{l,t}^{\rm super} (L) = \frac{\partial}{\partial t}\sum_n\frac{e^{-\beta E_n}}{\mathcal{Z}}\bigg(\langle n_I, t|\hat{H}_0+\hat{V}_I(t)+\hat{V}_{\rm dipole\,(I)}(t)+\hat{H}_I'(t)|n_I,t\rangle-\langle n|\hat{H}_0+\hat{V}+\hat{V}_{\rm dipole}|n\rangle\bigg)
\end{eqnarray}
\endwidetext
where $|n_I,t\rangle =\mathcal{T}e^{-\frac{i}{\hbar}\int_{-\infty}^t \hat{H}_I'(t')dt'}|n\rangle$ is the wave function in the interaction picture, and $\hat{H}'_I(t)$ and $\hat{V}_I(t)$ are interaction picture operators. For an arbitrary operator $\hat{A}$, the interaction picture operator is given by $\hat{A}_I(t)=e^{i\hat{H}_0t/\hbar}\hat{A}e^{-i\hat{H}_0t/\hbar}$.

Besides the energy absorption contributions we obtained in Eq.(\ref{24}), Eq.(\ref{25}) and Eq.(\ref{26}), there is one extra term which comes from the electric dipole-dipole interaction. Up to the second order of external phonon strain field $\bm{e}(\vec x, t)$, the contribution of phonon energy absorption from electric dipole-dipole interaction is given as follows,
\begin{eqnarray}\label{37}
\dot{ E}_{l}^{\rm dipole}
 & = & \frac{94A^2k^2N_0^3\ln N_0}{960\pi^2\epsilon^2}\left(1-e^{-\beta\hbar\omega}\right)\nonumber \\
 & {} & \omega\int {\,{\rm Im}\,\tilde{\chi}(T, \Omega)\,{\rm Im}\,\tilde{\chi}(T, \omega-\Omega)}d\Omega \nonumber \\
\dot{ E}_{t}^{\rm dipole}
 & = & \frac{53A^2k^2N_0^3\ln N_0}{960\pi^2\epsilon^2}\left(1-e^{-\beta\hbar\omega}\right)\nonumber \\
 & {} & \omega\int {\,{\rm Im}\,\tilde{\chi}(T, \Omega)\,{\rm Im}\,\tilde{\chi}(T, \omega-\Omega)}d\Omega
\end{eqnarray}
Let us discuss the length scale dependence of Eq.(\ref{37}). It is convenient to define the ``energy absorption rate per volume" $\dot{ \epsilon}_{l,t}^{\rm dipole}=(N_0L)^{-3}\dot{ E}_{l,t}^{\rm dipole}$. $\dot{ \epsilon}_{l,t}^{\rm dipole}$ has the same same length scale dependence as $\dot{ \epsilon}^{(2,3)}_{l,t}$. Therefore, the ratio between $\dot{ \epsilon}_{l,t}^{\rm dipole}$ and $\dot{ v}_{l,t}$ is given as follows,
\begin{eqnarray}\label{38}
\frac{\dot{ \epsilon}_{l,t}^{\rm dipole}}{\dot{ v}_{l,t}}\approx \frac{\frac{(\,{\rm Im}\,\tilde{\chi})^2}{\epsilon^2}\omega_c}{\frac{L^3(\,{\rm Im}\,\tilde{\chi}_t)^2}{\rho c_{l,t}^2}\ln(\omega_c/\omega)}
\end{eqnarray}
where in the above result we assume dielectric susceptibility $\,{\rm Im}\,\tilde{\chi}(T, \omega)\approx \,{\rm Im}\,\tilde{\chi}(T)$ is roughly independent of frequency within a certain range $\omega<\omega_c$. If we assume that there is a critical length scale $L_c$, beyond which $\dot{ \epsilon}_{l,t}^{\rm dipole}$ is smaller than $\dot{ v}_{l,t}$, then the critical length scale $L_c$ is 

\begin{eqnarray}
L_c=\left(\frac{({\rm Im}\,\tilde{\chi})^2\rho c_{l,t}^2\omega_c}{\epsilon^2({\rm Im}\,\tilde{\chi}_t)^2\ln(\omega_c/\omega)}\right)^{1/3}\sim 10\AA
\end{eqnarray}
where we let $\omega_c$ to take an extremely high value, $\omega_c\sim 10^{15}$rad/s. The above result indicates that the upper limit of $L_c$ is still smaller than the starting microscopic length scale of the renormalization procedure. Throughout the entire renormalization procedure the contribution of electric dipole-dipole interaction to resonance phonon energy absorption is always negligible compared to $\dot{v}_{l,t}$.

\section{Conclusion}
In this paper we develop a generic coupled block model to explain the universal properties of low-temperature glasses. We expand the general glass Hamiltonian in orders of long wavelength phonon strain field. The coefficient of the first order expansion is the non-elastic stress tensor $\hat{T}_{ij}^{(s)}$. The main assumption of our model is that the correlation function between stress tensors at different locations is diagonal in spacial coordinates: $\langle \hat{T}_{ij}^{(s)}\hat{T}_{kl}^{(s')}\rangle =\langle \hat{T}_{ij}^{(s)}\hat{T}_{kl}^{(s)}\rangle \delta_{ss'}$. We investigate the universal property of Meissner-Berret ratio in TTLS model for 13 different kinds of amorphous materials. The ratio $\gamma_l/\gamma_t$ equals to $1.06c_l/c_t$, independent of the materials' microscopic properties. The theoretical results are in fairly good agreement with the experimental data. We believe that such kind of universality essentially comes from the RKKY-type many-body interactions. 

We combine a set of single blocks to form a super block of glass, and allow the virtual phonons to exchange with each other. Such virtual phonon exchange process will generate non-elastic stress-stress interaction with the $1/r^{3}$ long range behavior. As the system size increases, more and more glass single blocks join in the glass total Hamiltonian. Eventually at experimental length scales, non-elastic stress-stress interaction dominates glass super block Hamiltonian. Starting from microscopic length scales, we use real space renormalization technique to carry out the Meissner-Berret ratio at experimental length scales.

External phonon strain fields have three ways to change non-elastic stress-stress interaction: it can modify the coefficient of non-elastic stress-stress interaction, $\Lambda_{ijkl}^{(ss')}$, by changing the relative positions of single blocks $\vec x_s-\vec x_s'$; it can change the stress tensor operators $\hat{T}_{ij}^{(s)}$; it perturbs the wave function $|n, t\rangle=\mathcal{T}e^{\frac{1}{i\hbar}\int_{-\infty}^t\hat{H}'(t')dt'}|n\rangle$. The contributions to resonance phonon energy absorption from the change of $\Lambda_{ijkl}^{(ss')}$ and $\hat{T}_{ij}^{(s)}$ decrease $\propto L^{-3}$ with the increase of length scales, while the energy absorption contribution from the change of wave function $|n, t\rangle$ is scale invariant. 

Among the Meissner-Berret ratios measured in 13 amorphous materials\cite{Berret1988}, 10 of them agree fairly good with our theoretical prediction. Other three materials, however, deviate more than 20\% compared to experimental data. We suggest such huge deviations come from the fact that the original experimental data of $\gamma_l/\gamma_t$ were not directly measured in experiment, instead they were inferred from TTLS model's parameters. According to the measurements by R. B. Stephens\cite{Stephens1973}, G. S. Cieloszyk, G. L. Salinger\cite{Stephens1972}, R. C. Zeller and R. O. Pohl\cite{Zeller1971}, the low temperature heat capacity of these 3 materials do not agree with TTLS model's expectation, which means TTLS model may not be a suitable description for them. Experimental data inferred from TTLS model may deviate from their original natures.

\section{Acknowledgement}
D.Z. would like to thank insightful discussions with Dervis C. Vural and Xianhao Xin. This work is supported by the National Science Foundation under Grant No. NSF-DMR 09-06921 at the University of Illinois.

\appendix
\widetext
\section{Derivation Details of Non-Elastic Stress-Stress Interaction Coefficient $\Lambda_{ijkl}^{(ss')}$ }
It was Joffrin and Levelut\cite{Joffrin1976} who first gave the detailed derivation of non-elastic stress-stress interaction $\hat{V}=\sum_{ijkl}\sum_{ss'}\Lambda_{ijkl}^{(ss')}\hat{T}_{ij}^{(s)}\hat{T}_{kl}^{(s')}$ in glasses. We give a further correction to their result here. To compare their result with ours, let us denote $(\Lambda_{ijkl}^{(ss')})_{\rm JL}$ for the coefficient of non-elastic stress-stress interaction derived by Joffrin and Levelut:
\begin{eqnarray}\label{A1}
 & {} & \hat{V} = \sum_{s\neq s'}^{N_0^3}\sum_{ijkl}(\Lambda_{ijkl}^{(ss')})_{\rm JL}\hat{T}_{ij}^{(s)}\hat{T}_{kl}^{(s')}\nonumber \\
 & {} & (\Lambda_{ijkl}^{(ss')})_{\rm JL} = -\frac{(\tilde{\Lambda}_{ijkl}(\vec n))_{\rm JL}}{8\pi\rho c_t^2|\vec x_s-\vec x_s'|^3}\nonumber \\
 & {} & (\tilde{\Lambda}_{ijkl}(\vec n))_{\rm JL} = -2(\delta_{jl}-3n_jn_l)\delta_{ik}\nonumber \\
 & {} &  +2\alpha\left\{-(\delta_{ij}\delta_{kl}+\delta_{ik}\delta_{jl}+\delta_{jk}\delta_{il})+3(n_in_j\delta_{kl}+n_in_k\delta_{jl}+n_in_l\delta_{jk}+n_jn_k\delta_{il}+n_jn_l\delta_{ik}+n_kn_l\delta_{ij})-15n_in_jn_kn_l\right\}\nonumber \\
\end{eqnarray}
where $\alpha=1-c_t^2/c_l^2$. We start our derivation from the glass Hamiltonian written as the summation of long wavelength phonon part, phonon strain-stress coupling and the purely non-elastic part of glass Hamiltonian:
\begin{eqnarray}\label{A2}
\hat{H}= \sum_{\vec q\mu}\left(\frac{| {p}_{\mu}(\vec q)|^2}{2m}+\frac{1}{2}m\omega^2_{\vec q\mu}| {u}_{\mu}(\vec q)|^2\right)+\sum_{s}\sum_{ij}e_{ij}^{(s)}\hat{T}_{ij}^{(s)}+\hat{H}_{0}^{\rm non}
\end{eqnarray}
where $\mu$ is the phonon polarization, i.e., longitudinal and transverse; $\vec q$ is the momentum and $m$ is the mass of elementary glass block. ${p}_{\mu}(\vec q)$ and $ {u}_{\mu}(\vec q)$ are momentum and displacement operators of phonon modes. Strain field $e_{ij}^{(s)}$ is defined the same as Eq.(\ref{2}), $e_{ij}^{(s)}=\frac{1}{2}({\partial u_{i}^{(s)}}/{\partial x_j}+{\partial u_{j}^{(s)}}/{\partial x_i})$. The relation of displacement operator $\vec u^{(s)}$ and $\vec u_{\mu}(\vec q)$ is set up by Fourier transformation:
\begin{eqnarray}\label{A3}
u_{i}^{(s)}=\frac{1}{\sqrt{N}}\sum_{\vec q\mu}u_{\mu}(\vec q){\rm e}_{\mu i}(\vec q)e^{i\vec q\cdot\vec x_s}
\end{eqnarray}
where $\vec {\rm e}_{\mu }(\vec q)$ is the unit vector representing the direction of vibrations, $N$ is the number density of glass elementary block. If we denote $a$ to be the length scale glass elementary block, then we have the relation $Nm/a^3=\rho$, where $\rho $ is the mass density of glass. For longitudinal mode $\mu=l$,
$
{\rm e}_{li}(\vec q)={q_i}/{q}
$, whereas for transverse modes $t_1$ and $t_2$, we have, 
\begin{eqnarray}\label{A4}
 & {} & \vec {\rm e}_{t_1}(\vec q)\cdot\vec q=\vec {\rm e}_{t_2}(\vec q)\cdot \vec q=\vec {\rm e}_{t_1}(\vec q)\cdot \vec {\rm e}_{t_1}(\vec q)=0\nonumber \\
 & {} & \sum_{\mu=t_1,t_2}{\rm e}_{\mu i}(\vec q){\rm e}_{\mu j}(\vec q)=\delta_{ij}-\frac{q_iq_j}{q^2}
\end{eqnarray}
the strain field is therefore written as 
$
e_{ij}^{(s)}=\frac{1}{2\sqrt{N}}\sum_{\vec q\mu}iu_{\mu}(\vec q)e^{i\vec q\cdot \vec x_s}\left[q_j{\rm e}_{\mu i}(\vec q)+q_{i}{\rm e}_{\mu j}(\vec q)\right]
$. Since for an arbitrary function $f(\vec q)$ we always have the following relation, 
$
\sum_{\vec q}f(\vec q)=\sum_{\vec q}\frac{1}{2}\left[f(\vec q)+f(-\vec q)\right]
$, and the displacement $u_{i}(\vec x)$ is real, i.e., $u_{i}(\vec x)=u^*_i(\vec x)$, we have $
u_{\mu i}(\vec q)=u_{\mu i}^*(-\vec q)
$.
With these properties of $u_{\mu}(\vec q)$ operators we can rewrite the stress-strain coupling term as follows, 
\begin{eqnarray}\label{A5}
\sum_{s}\sum_{ij}e_{ij}^{(s)}\hat{T}_{ij}^{(s)}
 & = & \frac{1}{4\sqrt{N}}\sum_{ij}\sum_{s}\sum_{\vec q\mu}\left[\left(iu_{\mu}(\vec q)e^{i\vec q\cdot \vec x_s}\right)+\left(iu_{\mu}(\vec q)e^{i\vec q\cdot \vec x_s}\right)^*\right](q_j{\rm e}_{\mu j}(\vec q)+q_j{\rm e}_{\mu i}(\vec q))\hat{T}_{ij}^{(s)}
\end{eqnarray}
Because the stress-strain coupling term is linear in displacement operators $u_{\mu}(\vec q)$, we can absorb it into terms quadratic in $u_{\mu}(\vec q)$, i.e., the quadratic displacement term of phonon Hamiltonian, by completing the square. An extra term comes out as follows:
\begin{eqnarray}\label{A6}
\hat{H}=\sum_{\vec q\mu}\left(\frac{|p_{\mu}(\vec q)|^2}{2m}+\frac{m\omega^2_{\vec q\mu}}{2}|u_{\mu}(\vec q)-u_{\mu}^{(0)}(\vec q)|^2-\frac{m\omega^2_{\vec q\mu}}{2}|u_{\mu}^{(0)}(\vec q)|^2\right)+\hat{H}^{\rm non}
\end{eqnarray}
where the ``equilibrium position" $u_{\mu}^{(0)}(\vec q)$ is 
\begin{eqnarray}\label{A7}
u_{\mu}^{(0)}(\vec q)=\frac{i}{2\sqrt{N}m\omega_{\vec q\mu}^2}\sum_{ij}\sum_{s}\bigg[q_j{\rm e}_{\mu i}(\vec q)+q_{i}{\rm e}_{\mu j}(\vec q)\bigg]\hat{T}_{ij}^{(s)}e^{-i\vec q\cdot \vec x_s}
\end{eqnarray}
After ``completing the square, $\frac{m\omega^2_{\vec q\mu}}{2}|u_{\mu}(\vec q)-u_{\mu}^{(0)}(\vec q)|^2$", the extra term $-\frac{m\omega^2_{\vec q\mu}}{2}|u_{\mu}^{(0)}(\vec q)|^2$ left out is the effective interaction between non-elastic stress tensors. It can be rewritten into two parts, the first part represents non-elastic stress-stress interaction within the same block, while the second part represents the interaction between blocks at different locations:
\begin{eqnarray}\label{A8}
 & {} & -\sum_{\vec q\mu}\left(\frac{m\omega^2_{\vec q\mu}}{2}|u_{\mu}^{(0)}(\vec q)|^2\right)\nonumber \\
 & = & -\sum_{\vec q\mu}\frac{1}{8Nm\omega^2_{\vec q\mu}}\sum_{ijkl}\bigg[q_{j}{\rm e}_{\mu i}(\vec q)+q_i{\rm e}_{\mu j}(\vec q)\bigg]
\bigg[q_{k}{\rm e}_{\mu l}(\vec q)+q_l{\rm e}_{\mu k}(\vec q)\bigg]\sum_{s}\hat{T}_{ij}^{(s)}\hat{T}_{kl}^{(s)}\nonumber \\
 & {} & -\sum_{\vec q\mu}\frac{1}{8Nm\omega^2_{\vec q\mu}}\sum_{ijkl}\bigg[q_{j}{\rm e}_{\mu i}(\vec q)+q_i{\rm e}_{\mu j}(\vec q)\bigg]
\bigg[q_{k}{\rm e}_{\mu l}(\vec q)+q_l{\rm e}_{\mu k}(\vec q)\bigg]\sum_{s\neq s'}\hat{T}_{ij}^{(s)}\hat{T}_{kl}^{(s')} \cos(\vec q\cdot (\vec x_{s}-\vec x_{s}'))
\end{eqnarray}
We denote the second term in Eq.(\ref{A8}) as non-elastic stress-stress interaction $\hat{V}$. We apply the properties of unit vector for longitudinal and transverse phonons, it can be further simplified as 
\begin{eqnarray}\label{A9}
\hat{V} 
 & = & \frac{1}{2Nm}\left(\frac{1}{c_t^2}-\frac{1}{c_l^2}\right)\sum_{s\neq s'}\sum_{ijkl}\sum_{\vec q}\left(\frac{q_iq_jq_kq_l}{q^4}\right)\cos(\vec q\cdot\vec x_{ss'})T_{ij}^{(s)}T_{kl}^{(s')}\nonumber \\
 & {} & -\frac{1}{8Nm}\frac{1}{c_t^2}\sum_{s\neq s'}\sum_{ijkl}\sum_{\vec q}\left(\frac{q_jq_l\delta_{ik}+q_jq_k\delta_{il}
+q_iq_l\delta_{jk}+q_iq_k\delta_{jl}
}{q^2}\right)\cos(\vec q\cdot \vec x_{ss'})T_{ij}^{(s)}T_{kl}^{(s')}
\end{eqnarray}
where $\vec x_{ss'}=\vec x_s-\vec x_s'$. If we assume that the length scale of elementary block $a$ is much smaller than phonon wavelength (long wavelength limit), we can replace the summation over momentum $\vec q$ with momentum integral. For simplicity we write $\hat{V}$ into two parts, $\hat{V}^{(1)}$ and $\hat{V}^{(2)}$ as follows:
\begin{eqnarray}\label{A10}
\hat{V}^{(1)} & = & \frac{a^3}{2Nm}\left(\frac{1}{c_t^2}-\frac{1}{c_l^2}\right)\sum_{s\neq s'}\sum_{ijkl}\left\{\int\frac{d^3q}{(2\pi)^3}\left(\frac{q_iq_jq_kq_l}{q^4}\right)\cos(\vec q\cdot \vec x_{ss'})\right\}\hat{T}_{ij}^{(s)}\hat{T}_{kl}^{(s')}\nonumber \\
\hat{V}^{(2)} & = & -\frac{a^3}{8Nm}\frac{1}{c_t^2}\sum_{s\neq s'}\sum_{ijkl}\left\{\int\frac{d^3q}{(2\pi)^3}\left(\frac{q_jq_l\delta_{ik}+q_jq_k\delta_{il}
+q_iq_l\delta_{jk}+q_iq_k\delta_{jl}
}{q^2}\right)\cos(\vec q\cdot \vec x_{ss'})\right\}\hat{T}_{ij}^{(s)}\hat{T}_{kl}^{(s')}
\end{eqnarray}
In the above two equations, we need to evaluate the following two integrals:
\begin{eqnarray}\label{A11}
f^{(1)}_{ijkl} & = & \int \frac{d^3q}{(2\pi)^3}\frac{q_iq_jq_kq_l}{q^4}\cos(\vec q\cdot \vec x)\nonumber \\
f^{(2)}_{jl} & = & \int \frac{d^3q}{(2\pi)^3}\frac{q_jq_l}{q^2}\cos(\vec q\cdot \vec x)
\end{eqnarray}
Let us introduce a new parameter $\lambda$ and take the limit $\lambda\to 0$:
\begin{eqnarray}\label{A12}
f^{(1)}_{ijkl}(\lambda) & = & \left(\frac{\partial}{\partial x_i}\frac{\partial}{\partial x_j}\frac{\partial}{\partial x_k}\frac{\partial}{\partial x_l}\right)\int \frac{d^3q}{(2\pi)^3}\frac{1}{(q^2+\lambda^2)^2}\frac{1}{2}\left(e^{i\vec q\cdot \vec x}+e^{-i\vec q\cdot \vec x}\right)\nonumber \\
f^{(2)}_{jl}(\lambda) & = & -\left(\frac{\partial}{\partial x_j}\frac{\partial }{\partial x_l}\right)\int \frac{d^3q}{(2\pi)^3}\frac{1}{(q^2+\lambda^2)}\frac{1}{2}\left(e^{i\vec q\cdot \vec x}+e^{-i\vec q\cdot \vec x}\right)
\end{eqnarray}
Using contour integral, and choose the pole at $q=-i\lambda$, we have, 
\begin{eqnarray}\label{A13}
f^{(1)}_{ijkl}(\lambda) & = & \left(\frac{\partial}{\partial x_i}\frac{\partial}{\partial x_j}\frac{\partial}{\partial x_k}\frac{\partial}{\partial x_l}\right)\frac{1}{8\pi\lambda}e^{-\lambda x}\nonumber \\
f^{(2)}_{jl}(\lambda) & = & -\left(\frac{\partial}{\partial x_j}\frac{\partial }{\partial x_l}\right)\frac{1}{4\pi x}e^{-\lambda x}
\end{eqnarray}
Taking the derivatives in Eq.(A13), we obtain the following results, 
\begin{eqnarray}\label{A14}
\lim_{\lambda\to 0}f_{ijkl}^{(1)}(\lambda) & = & \frac{1}{8\pi x^3}\bigg\{(\delta_{ij}\delta_{kl}+\delta_{ik}\delta_{jl}+\delta_{jk}\delta_{il})\nonumber \\
 & {} & \quad\quad\quad-3(n_in_j\delta_{kl}+n_in_k\delta_{jl}+n_in_l\delta_{jk}+n_jn_k\delta_{il}+n_jn_l\delta_{ik}+n_kn_l\delta_{ij})+15n_in_jn_kn_l\bigg\}\nonumber \\
\lim_{\lambda\to 0}f_{jl}^{(2)}(\lambda) & = & \frac{1}{4\pi x^3}\left(\delta_{jl}-3n_jn_l\right)
\end{eqnarray}
Finally, plugging the above results of integrals, we eventually get our non-elastic stress-stress interaction coefficient $\Lambda_{ijkl}^{(ss')}$ as follows, 
\begin{eqnarray}\label{A15}
 & {} & \hat{V}=\sum_{s\neq s'}\sum_{ijkl}\Lambda_{ijkl}^{(ss')}\hat{T}_{ij}^{(s)}\hat{T}_{kl}^{(s')}\nonumber \\
 & {} & \Lambda_{ijkl}^{(ss')} = -\frac{\tilde{\Lambda}_{ijkl}(\vec n)}{8\pi \rho c_t^2|x_s-x_s'|^3}\nonumber \\
 & {} & \tilde{\Lambda}_{ijkl}(\vec n) = \frac{1}{4}\left\{(\delta_{jl}-3n_jn_l)\delta_{ik}+(\delta_{jk}-3n_jn_k)\delta_{il}+(\delta_{ik}-3n_in_k)\delta_{jl}
+(\delta_{il}-3n_in_l)\delta_{jk}\right\}\nonumber \\
 & {} & +\frac{1}{2}\alpha\left\{-(\delta_{ij}\delta_{kl}+\delta_{ik}\delta_{jl}+\delta_{jk}\delta_{il})+3(n_in_j\delta_{kl}+n_in_k\delta_{jl}+n_in_l\delta_{jk}+n_jn_k\delta_{il}+n_jn_l\delta_{ik}+n_kn_l\delta_{ij})-15n_in_jn_kn_l\right\}\nonumber \\
\end{eqnarray}
Compare $\Lambda_{ijkl}^{(ss')}$ in Eq.(\ref{A15}) with $(\Lambda_{ijkl}^{(ss')})_{\rm JL}$ in Eq.(\ref{A1}), there are 4 differences between our result and the result by Joffrin and Levelut. To illustrate these differences, we separate $\Lambda_{ijkl}^{(ss')}$ and $(\Lambda_{ijkl}^{(ss')})_{\rm JL}$ into two parts as follows:
\begin{eqnarray}\label{A16}
 & {} & \tilde{\Lambda}_{ijkl}(\vec n) = \tilde{\Lambda}_{ijkl}^{(1)}(\vec n)+\tilde{\Lambda}^{(2)}_{ijkl}(\vec n)\nonumber \\
 & {} & \tilde{\Lambda}^{(1)}_{ijkl}(\vec n) = \frac{1}{4}\left\{(\delta_{jl}-3n_jn_l)\delta_{ik}+(\delta_{jk}-3n_jn_k)\delta_{il}+(\delta_{ik}-3n_in_k)\delta_{jl}
+(\delta_{il}-3n_in_l)\delta_{jk}\right\}\nonumber \\
 & {} & \tilde{\Lambda}^{(2)}_{ijkl}(\vec n) = \frac{1}{2}\alpha\left\{-(\delta_{ij}\delta_{kl}+\delta_{ik}\delta_{jl}+\delta_{jk}\delta_{il})+3(n_in_j\delta_{kl}+n_in_k\delta_{jl}+n_in_l\delta_{jk}+n_jn_k\delta_{il}+n_jn_l\delta_{ik}+n_kn_l\delta_{ij})-15n_in_jn_kn_l\right\}\nonumber \\\nonumber \\
 & {} & (\tilde{\Lambda}_{ijkl}(\vec n))_{\rm JL} =(\tilde{\Lambda}^{(1)}_{ijkl}(\vec n))_{\rm JL} +(\tilde{\Lambda}^{(2)}_{ijkl}(\vec n))_{\rm JL} \nonumber \\
 & {} & (\tilde{\Lambda}_{ijkl}^{(1)}(\vec n))_{\rm JL} = -2(\delta_{jl}-3n_jn_l)\delta_{ik}\nonumber \\
 & {} & (\tilde{\Lambda}_{ijkl}^{(2)}(\vec n))_{\rm JL} = 2\alpha\left\{-(\delta_{ij}\delta_{kl}+\delta_{ik}\delta_{jl}+\delta_{jk}\delta_{il})+3(n_in_j\delta_{kl}+n_in_k\delta_{jl}+n_in_l\delta_{jk}+n_jn_k\delta_{il}+n_jn_l\delta_{ik}+n_kn_l\delta_{ij})-15n_in_jn_kn_l\right\}\nonumber \\
\end{eqnarray}

(1). $\tilde{\Lambda}^{(1)}_{ijkl}(\vec n)$ and $(\tilde{\Lambda}^{(1)}_{ijkl}(\vec n) )_{\rm JL}$ are different. This difference is not a big problem, because in the derivation by Joffrin and Levelut, their phonon strain field $e_{ij}$ is defined as $e_{ij}=\partial u_i/\partial x_j$, while in our derivation, we define phonon strain field $e_{ij}=\frac{1}{2}(\partial u_i/\partial x_j+\partial u_j/\partial x_i)$. According to D. C. Vural and A. J. Leggett\cite{Leggett2011}, because the anti-symmetric part of the tensor $\partial u_j/\partial x_i$ corresponds to a local rotation, and a spacially uniform rotation costs no energy, we drop the anti-symmetric part of phonon strain field for simplicity. \\

(2). The coefficient $\tilde{\Lambda}_{ijkl}(\vec n)$ derived by us is smaller by a total factor of $1/2$ compared to the coefficient $(\tilde{\Lambda}_{ijkl}(\vec n))_{\rm JL}$ derived by Joffrin and Levelut; \\

(3). The term $\tilde{\Lambda}^{(1)}_{ijkl}(\vec n)$ derived by us does not have a negative sign in front of it: $\frac{1}{4}\{(\delta_{jl}-3n_jn_l)\delta_{ik}+(\delta_{jk}-3n_jn_k)\delta_{il}+(\delta_{ik}-3n_in_k)\delta_{jl}
+(\delta_{il}-3n_in_l)\delta_{jk}\}$, while the term $(\tilde{\Lambda}^{(1)}_{ijkl}(\vec n))_{\rm JL}$ by Joffrin and Levelut has: $-2(\delta_{jl}-3n_jn_l)\delta_{ik}$;\\

(4). The term $ \tilde{\Lambda}^{(2)}_{ijkl}(\vec n)$
derived by us has an extra factor of $1/2$ compared to $ (\tilde{\Lambda}^{(2)}_{ijkl}(\vec n))_{\rm JL}$.

\endwidetext

\end{document}